\documentclass[12pt]{article}
\input{tcilatex}

\begin{document}

\pagestyle{empty}

\noindent {\small USC-98/HEP-B3\hfill \hfill hep-th/9806085}\newline
{\small \hfill }

{\vskip 0.7cm}

\begin{center}
{\Large Gauge Symmetry in Phase Space with Spin,}

{\Large A Basis for Conformal Symmetry and Duality }

{\Large Among Many Interactions}$^{*}$ \\[0pt]

{\vskip 0.5cm}

\textbf{Itzhak Bars and Cemsinan Deliduman} {\Large \ \\[0pt]
}

{\vskip 0.4cm}

\textbf{Department of Physics and Astronomy}

\textbf{University of Southern California}

\textbf{Los Angeles, CA 90089-0484}

{\vskip 0.5cm}

\textbf{ABSTRACT}
\end{center}

\noindent We show that a simple OSp$\left( 1/2\right) $ worldline gauge
theory in 0-brane phase space $X^{M}\left( \tau \right) ,P^{M}\left( \tau
\right) $ with spin degrees of freedom $\psi ^{M}\left( \tau \right) $,
formulated for a $d+2$ dimensional spacetime with two times $X^{0}\left(
\tau \right) ,X^{0^{\prime }}\left( \tau \right) $, unifies many physical
systems which ordinarily are described by a 1-time formulation. Different
systems of 1-time physics emerge by choosing gauges that embed ordinary time
in $d+2$ dimensions in different ways. The embeddings have different
topology and geometry for the choice of time among the $d+2$ dimensions.
Thus, 2-time physics unifies a large number of 1-time physical interacting
systems, and establishes a kind of duality among them. One manifestation of
the two times is that all of these physical systems have the same quantum
Hilbert space in the form of a unique representation of SO$\left( d,2\right) 
$ with the same Casimir eigenvalues. By changing the number of spinning
degrees of freedom $\psi _{a}^{M}\left( \tau \right) $, $a=1,2,\cdots ,n$,
(including no spin $n=0)$, the gauge group changes to OSp$\left( n/2\right) $%
. Then the eigenvalue of the Casimirs of SO$\left( d,2\right) $ depend on $n$
and the content of the 1-time physical systems that are unified in the same
representation depend on $n$. The models we study raise new questions about
the nature of spacetime.

\vfill \hrule width 6.7cm \vskip 2mm

{$^{*}$ {\small Research partially supported by the US. Department of Energy
under grant number DE-FG03-84ER40168.}}

\vfill\eject

\setcounter{page}1\pagestyle{plain}

\section{Introduction}

In two recent papers \cite{dualconf}\cite{dualH} we showed that various
physical systems, which normally are considered unrelated, are actually
unified by the same theory that establishes a kind of duality among them.
Examples of such systems included the free relativistic particle in $d$
spacetime dimensions, the H-atom in $d-1$ space dimensions and the harmonic
oscillator in $d-2$ space dimensions with its mass identified with a
momentum in an extra dimension. Related ideas were considered in \cite%
{marnelius}-\cite{siegel}. Our aim in this paper is twofold. First, to
generalize the theory to describe spinning systems, and second to present an
infinite array of interacting models (relativistic, non-relativistic,
arbitrary potentials, curved backgrounds, etc.) with spin, that are unified
by gauge transformations (dualities) in the same theory. Our understanding
of the quantum version of the theory is solidified by working out many
examples and gauge choices in detail.

Through all these examples we emphasize that 1-time physics with various
interactions are unified in some geometrical sense as 2-time physics. The
theory in \cite{dualconf}\cite{dualH} is a simple Sp$\left( 2\right) $ gauge
theory on the worldline. Sp$\left( 2\right) $ is the global isometry group
of the quantum relations $\left[ x,p\right] =i$ and it transforms ($x,p)$ as
a doublet. The idea was to turn this group into a local symmetry of some
theory. This was loosely motivated by the fact that all known dualities
involve a transformation of canonically conjugate phase space variables. The
Sp$\left( 2\right) $ gauge theory on the worldline achieved this, but it
required that the worldline vectors $\left( X^M\left( \tau \right)
,P^M\left( \tau \right) \right) $ be in a spacetime with two timelike
coordinates $X^0\left( \tau \right) ,X^{0^{\prime }}\left( \tau \right) $.
This turned out to be a boon rather than a drawback. The presence of two
times together with the larger gauge symmetry allowed the possibility of
gauge choices that are ghost free and physical (unitary). The gauge fixed
theory has a single time. The ability to choose time in various ways turned
out to be equivalent to different choices of Hamiltonians that describe
ordinary 1-time physics. In this way, different looking physics corresponds
to gauge choices within the same theory. The gauge transformations that map
them into each other may be interpreted as dualities (in a universe of two
times).

In this paper the theory is generalized by including anticommuting phase
space variables for worldline fermions $\psi ^M\left( \tau \right) $. The
gauge group becomes OSp$\left( 1/2\right) $, and $(\psi ^M,X^M,P^M)$ form a
triplet. At the end of the paper we further generalize this to $n$ worldline
fermions $\psi _a^M\left( \tau \right) $, $a=1,2,\cdots ,n$, and gauge group
OSp$\left( n/2\right) $. The requirement that this be a 2-time theory
remains the same for any $n$. The content of the 1-time physical dual
sectors changes as a function of $n$. However, for a fixed $n$ all dual
physics is described within the same quantum Hilbert space that corresponds
to a unique unitary representation of SO$\left( d,2\right) $ with fixed
Casimir eigenvalues.

The paper is organized as follows. After formulating the OSp$\left(
1/2\right) $ gauge theory, we quantize it covariantly, and show that the
gauge invariant states must be described by a unique representation of SO$%
\left( d,2\right) $ with fixed Casimir eigenvalues. Next we choose specific
gauges, which we call ``particle gauge'', ``light-cone gauge'', ``H-atom
gauge'', ``AdS gauge'', ``Conformal gauge'', and study the quantum theory in
each of those gauges. We show that the physics looks different according to
the gauge choice of time, but that the Hilbert space is the same in each
case, and that it has the same eigenvalues of the Casimirs of SO$\left(
d,2\right) $. In a semi-classical approach in the ``H-atom gauge'' we show
that any Hamiltonian of the form $H=\mathbf{p}^2/2+V\left( \mathbf{r,p,S}%
\right) $ with any potential energy function $V,$ emerges as a gauge choice.
At the end of the paper we argue that when $n$ changes, the spin content
changes. For example in the particle gauge the relativistic particle that is
described corresponds to the antisymmetric form $A_{\mu _1\mu _2\cdots \mu
_{p+1}}\left( x\right) $ that couples to $p$-branes, with $p=n/2-1$ for even 
$n$, and similar fermionic counterparts for odd $n$.

The message of our work is that 2-time physics is not only possible, but
also is a basis for unifying many features of 1-time physics in some
geometrical manner. This raises new questions about the nature of time and
space. Our gauge symmetry approach in 0-brane phase space connects together
dualities and 2-time physics inextricably from each other, and gives a new
rich area to explore further and generalize to higher p-branes. Our work
supports the idea that the fundamental theory of our universe may be better
understood in a 2-time formulation, as various hints and theories have
suggested from different directions \cite{duff}-\cite{sezrudy}.

\section{Gauging OSp$\left( 1/2\right) $}

OSp$\left( 1/2\right) $ has two local fermionic $s^i\left( \tau \right) $
and three local bosonic $\omega ^{ij}\left( \tau \right) $ parameters. Under
the subgroup Sp$\left( 2,R\right) $ the $s^i$ with $i=1,2$ form a doublet,
while the symmetric $\omega ^{ij}=\omega ^{ji}$ form a triplet. Consider the
OSp$\left( 1/2\right) $ triplets $\Phi _a^M\left( \tau \right) =\left( \psi
^M,X_1^M,X_2^M\right) $ (one for each $M$) which transform like the
fundamental representation of OSp$\left( 1/2\right) $%
\begin{equation}
\delta \psi ^M=s^iX_i^M,\quad \delta X_i^M=\varepsilon _{ik}\left( \omega
^{kl}X_l^M-is^k\psi ^M\right) .  \label{osp12}
\end{equation}
The complex number $i$ is introduced in $\delta X_i^M$ to insure that the
product of fermions $is^k\psi ^M$ is hermitian, assuming that each fermion
is hermitian individually. For each $M$, $\psi ^M$ is a singlet of Sp$\left(
2,R\right) $ while $X_i^M$ is a doublet of Sp$\left( 2\right) $. Two such
triplets $\Phi _a^M,\Phi _b^N$ form an OSp$\left( 1/2\right) $ invariant $%
I^{MN}$ under the dot product with the metric $g^{ab}$ given by 
\begin{equation}
I^{MN}=\Phi _a^Mg^{ab}\Phi _b^N=X_i^M\varepsilon ^{ij}X_j^N-i\psi ^M\psi ^N.
\label{invariants}
\end{equation}

Fermionic and bosonic gauge potentials $\left( F^i,A^{ij}\right) $ are
introduced in one to one correspondance with the parameters. There are two
fermions $F^i\left( \tau \right) $ and three bosons $A^{ij}\left( \tau
\right) =A^{ji}\left( \tau \right) $. They transform as 
\begin{eqnarray}
\delta F^i &=&\partial _\tau s^i+\omega ^{ik}\varepsilon
_{kl}F^l-A^{ik}\varepsilon _{kl}s^l\,, \\
\delta A^{ij} &=&\partial _\tau \omega ^{ij}+\omega ^{ik}\varepsilon
_{kl}A^{lj}+\omega ^{jk}\varepsilon _{kl}A^{il}-is^iF^j-is^jF^i\,.
\end{eqnarray}
The following covariant derivatives $D_\tau \Phi _a^M=\left( D_\tau \psi
^M,D_\tau X_i^M\right) $ 
\begin{eqnarray}
D_\tau \psi ^M &=&\partial _\tau \psi ^M-F^iX_i^M\,, \\
D_\tau X_i^M &=&\partial _\tau X_i^M-\varepsilon _{ik}\left(
A^{kl}X_l^M-iF^k\psi ^M\right) \,,
\end{eqnarray}
transform like OSp$\left( 1/2\right) $ triplets 
\begin{eqnarray}
\delta \left( D_\tau \psi ^M\right) &=&s^iD_\tau X_i^M\,. \\
\delta \left( D_\tau X_i^M\right) &=&\varepsilon _{ik}\left( \omega
^{kl}D_\tau X_l^M-is^kD_\tau \psi ^M\right) \,.
\end{eqnarray}
We can then construct more OSp$\left( 1/2\right) $ invariants by using the
covariant derivatives and the metric defined in (\ref{invariants}) $\left(
D_\tau \Phi _a^M\right) g^{ab}\Phi _b^N$. In particular we construct a gauge
invariant action 
\begin{eqnarray}
S_0 &=&\frac 12\int_0^Td\tau \,\,\left( D_\tau \Phi _a^M\right) g^{ab}\Phi
_b^N\,\,\eta _{MN} \\
&=&\frac 12\int_0^Td\tau \,\,\left[ D_\tau X_i^M\varepsilon
^{ij}X_j^N-iD_\tau \psi ^M\psi ^N\,\,\right] \eta _{MN} \\
&=&\int_0^Td\tau \,\,\left[ X_2\cdot \partial _\tau X_1+\frac i2\psi \cdot
\partial _\tau \psi -\frac 12A^{ij}X_i\cdot X_j+iF^iX_i\cdot \psi \right] \,.
\label{actionS}
\end{eqnarray}
The equation of motion of the gauge fields give the following constraints 
\begin{equation}
X_i\cdot X_j=0,\quad X_i\cdot \psi =0.  \label{firstconstr}
\end{equation}
As in the purely bosonic case the signature of the metric $\eta ^{MN}$ must
be $\left( d,2\right) $ including two timelike dimensions otherwise the
constraints have no non-trivial solutions. The action is manifestly
invariant under SO$\left( d,2\right) $ transformations since the metric $%
\eta ^{MN}$ is invariant. The conserved generators of the symmetry have an
orbital part $L^{MN}$ and spin part $S^{MN}$ 
\begin{eqnarray}
J^{MN} &=&L^{MN}+S^{MN}\,,  \label{confgen} \\
L^{MN} &=&X_1^MX_2^N-X_1^NX_2^M\,, \\
S^{MN} &=&\frac 1{2i}\left( \psi ^M\psi ^N-\psi ^N\psi ^M\right) .
\end{eqnarray}
The total generators $J^{MN}$ are OSp$\left( 1/2\right) $ gauge invariant
according to (\ref{invariants}) (take the antisymmetric part of $I^{MN}$).

From the action we obtain the canonical conjugate pairs $X_1^M=X^M$ and $%
X_2^M=P^M$. Furthermore, the canonical conjugate to $\psi ^M$ is naively $%
i\psi ^M/2$, however this is also a second class constraint. Once the second
class constraint is taken into account, the commutation rules for quantizing
the system covariantly are 
\begin{equation}
\left[ X^M,P^N\right] =i\eta ^{MN},\quad \left\{ \psi ^M,\psi ^N\right\}
=\,\eta ^{MN}.
\end{equation}
The $\psi ^M$ form a Clifford algebra which is represented by gamma matrices 
$\psi ^M=\gamma ^M/\sqrt{2}$, where the gamma matrices are normalized in the
standard way $\left\{ \gamma ^M,\gamma ^N\right\} =2\,\eta ^{MN}$. The
quantum system is subject to first class constraints (\ref{firstconstr}) 
\begin{equation}
X\cdot X=P\cdot P=X\cdot P=X\cdot \psi =P\cdot \psi =0\,,
\label{constraints}
\end{equation}
which will be imposed on the Hilbert space. These constraints form the OSp$%
\left( 1/2\right) $ superalgebra defined by three bosonic and two fermionic
generators 
\begin{eqnarray}
J_3 &=&\frac 14\left( X^2+P^2\right) ,\quad J_1=\frac 14\left( X\cdot
P+P\cdot X\right) , \\
J_2 &=&\frac 14\left( X^2-P^2\right) ,\quad S_{\pm }=\frac 1{2\sqrt{2}%
}\left( P\pm iX\right) \cdot \psi . \\
J_{\pm } &=&\pm \frac 1{4i}\left( P\pm iX\right) ^2=J_1\pm iJ_2
\end{eqnarray}
The OSp$\left( 1/2\right) $ superalgebra among these first class constraints
is given by 
\begin{eqnarray}
\left[ J_3,J_1\right] &=&iJ_2,\quad \left[ J_3,J_2\right] =-iJ_1,\quad \left[
J_1,J_2\right] =-iJ_3 \\
\left[ J_3,S_{\pm }\right] &=&\pm \frac 12S_{\pm },\quad \left[ J_1,S_{\pm }%
\right] =\frac i2S_{\mp },\quad \left[ J_2,S_{\pm }\right] =\mp \frac
12S_{\mp } \\
\left\{ S_{\pm },S_{\pm }\right\} &=&\pm \frac i2\left( J_1\pm iJ_2\right)
,\quad \left\{ S_{+},S_{-}\right\} =\frac{J_3}2.
\end{eqnarray}
where $J_{\pm }=\left( J_1\pm iJ_2\right) $. Thus $\left( J_1,J_2,J_3\right) 
$ are represented on the row $\left( S_{+},S_{-}\right) $ $\,$by the Pauli
matrices $\left( \Sigma ^i\right) _{\,\,\,\alpha }^\beta =\left( i\sigma
_1/2,i\sigma _2/2,\sigma _3/2\right) $ respectively, i.e. $\left[
J_i,S_\alpha \right] =iS_\beta \,\left( \Sigma _i\right) _{\,\,\,\alpha
}^\beta $ and the last line may be written as $\left\{ S_\alpha ,S_\beta
\right\} =\Sigma _{\alpha \beta }^iJ_i$.

The quadratic and cubic Casimirs of the superalgebra OSp$\left( 1/2\right) $
are 
\begin{eqnarray}
C_2\left( OSp(1/2)\right) &=&J_3^2-J_1^2-J_2^2-S_{+}S_{-}+S_{-}S_{+}\,, \\
C_3\left( OSp(1/2)\right) &=&S_{+}J_3S_{-}+S_{-}J_3S_{+}+iS_{+}\left(
J_1-iJ_2\right) S_{+}-iS_{-}\left( J_1+iJ_2\right) S_{-}\,.  \nonumber
\end{eqnarray}
These commute with all the generators\thinspace $J_i,S_\alpha $. In terms of
the canonical operators, with the orders of operators taken into account,
the quadratic Casimir becomes 
\begin{eqnarray}
C_2\left( OSp(1/2)\right) &=&\frac 14\left( X^MP^2X_M-X\cdot P\,\,P\cdot
X\right) +\frac 1{16}\left( d^2-4\right) \\
&&+\frac 14\left( \frac 1{2i}\left[ \psi _M,\psi _N\right] \left(
X^MP^N-X^NP^M\right) \right) +\frac 14\psi \cdot \psi  \nonumber
\end{eqnarray}
Similarly, the cubic Casimir is computed in terms of the canonical operators.

Next, consider the quadratic Casimir operator of SO$\left( d,2\right) $
given by

\begin{eqnarray}
C_2\left( SO\left( d,2\right) \right) &=&\frac 12J^{MN}J_{MN}=\frac
12L^{MN}L_{MN}+\frac 12S^{MN}S_{MN}+L^{MN}S_{MN}  \nonumber \\
&=&\left( X^MP^2X_M-X\cdot P\,\,P\cdot X\right) +\frac 14\left( 2\left( \psi
\cdot \psi \right) ^2-\psi \cdot \psi \right)  \nonumber \\
&&+\frac 1{2i}\left[ \psi _M,\psi _N\right] \left( X^MP^N-X^NP^M\right) \,.
\end{eqnarray}
Therefore, we find the following relation between the quadratic Casimirs of
SO$\left( d,2\right) $ and OSp$\left( 1/2\right) $ 
\begin{equation}
C_2\left( SO\left( d,2\right) \right) =4C_2\left( OSp(1/2)\right) -\frac
18\left( d+2\right) \left( d-1\right) \,.
\end{equation}
Where we have used $\psi \cdot \psi =\frac 12\left( d+2\right) .$ Similarly,
the higher order Casimirs of the conformal group $C_n=\frac 1{n!}Tr\left(
iJ\right) ^n$ are obtained in terms of the Casimir operators of OSp$\left(
1/2\right) .$

In the gauge invariant sector the quadratic Casimir of the gauge group must
vanish $C_2\left( 1/2\right) =0$. Therefore, the physical sector is
characterized by 
\begin{equation}
C_2\left( OSp(1/2)\right) =0,\quad C_2\left( SO(d,2)\right) =-\frac 18\left(
d+2\right) \left( d-1\right) .  \label{c2sod2}
\end{equation}
Similarly, the eigenvalues $C_n\left( SO(d,2)\right) $ are completely fixed
after setting all OSp$\left( 1/2\right) $ Casimirs equal to zero. We will
not use the higher Casimirs in this paper. We will verify the result for $%
C_2\left( SO(d,2)\right) $ in non-covariant quantization in several gauges.

\section{Particle gauge \& Dirac equation}

Consider the basis $X^M=\left[ X^{+^{\prime }},X^{-^{\prime }},X^\mu \right] 
$ with non-zero metric components $\eta ^{+^{\prime }-^{\prime }}=\eta
^{-^{\prime }+^{\prime }}=-1$ and $\eta ^{\mu \nu }=diag\left( -1,+1,\cdots
,+1\right) $ Minkowski metric. Choose two bosonic gauges $X^{+^{\prime
}}=1,\,P^{+^{\prime }}=0$ and one fermionic gauge $\psi ^{+^{\prime }}=0$,
and solve explicitly two bosonic and one fermionic constraints $X^2=X\cdot
P=X.\psi =0$. We will call this the relativistic particle gauge. The
remaining degrees of freedom $x^\mu ,p^\mu ,\psi ^\mu $ are in Minkowski
spacetime and they parametrize $X^M,P^M,\psi ^M$ as follows

\begin{eqnarray}
M &=&\left[ \quad +^{\prime }\quad ,\quad -^{\prime }\quad ,\quad \mu \quad %
\right] \,,  \nonumber \\
X^M &=&\left[ \quad 1\quad ,\quad x^2/2\quad ,\quad x^\mu \quad \right] \,,
\label{relativistic} \\
P^M &=&\left[ \quad 0\quad ,\quad x\cdot p\quad ,\quad p^\mu \quad \right]
\,,\quad p^2=0\,,  \nonumber \\
\psi ^M &=&\left[ \quad 0\quad ,\quad x\cdot \psi \quad ,\quad \psi ^\mu
\quad \right] \,,\,\,\,\psi ^\mu p_\mu =0.  \nonumber
\end{eqnarray}
There is manifest SO$\left( d-1,1\right) $ Lorentz symmetry. There remains
one bosonic and one fermionic gauge degrees of freedom and the corresponding
constraints $p^2=0,\,\psi ^\mu p_\mu =0$. The quantum rules are $\left[
x^\mu ,p^\nu \right] =i\eta ^{\mu \nu }$, and $\left\{ \psi ^\mu ,\psi ^\nu
\right\} =\eta ^{\mu \nu }$. The quantum states are labelled by $|\alpha ,p>$%
, or $|\alpha ,x>$ with $p^2=0$ and $\psi ^\mu p_\mu =0$ to be satisfied on
states. The index $\alpha $ is a spinor index in $d$ dimensions, and $\psi
^\mu $ acts like the Dirac gamma matrix $\psi ^\mu \rightarrow \gamma ^\mu /%
\sqrt{2}$ on these states. Note that $\left( \sqrt{2}\psi ^\mu p_\mu \right)
^2=p^2$, so that the constraint $p^2=0$ need not be considered separately.

For the general physical state $|\Psi >$ the constraint $\sqrt{2}\psi ^\mu
p_\mu |\Psi >=0$ becomes the Dirac equation for a massless particle. When
expressed in $x$-space $<x,\alpha |\Psi >=\Psi _\alpha \left( x\right) $ the
physical state constraint takes the form of the Dirac equation 
\begin{equation}
<x,\alpha |\left( \sqrt{2}\psi ^\mu p_\mu \right) |\Psi >=-i\left( \gamma
^\mu \partial _\mu \Psi \right) _\alpha =0.
\end{equation}
The effective field theory is therefore given by the action for the free
Dirac field in $d$ dimensions 
\begin{equation}
S_{eff}=\int d^dx\,\overline{\Psi }i\gamma \cdot \partial \Psi .
\end{equation}

The conformal generators (\ref{confgen}) in this gauge take the form 
\begin{eqnarray}
J^{+^{\prime }-^{\prime }} &=&\frac 12\left( x\cdot p+p\cdot x\right) +is_0
\\
J^{+^{\prime }\mu } &=&p^\mu ,\quad J^{\mu \nu }=x^\mu p^\nu -x^\nu p^\mu
+s^{\mu \nu }  \label{ConfLorentz} \\
J^{-^{\prime }\mu } &=&\frac 12x_\lambda p^\mu x^\lambda -\frac 12x^\mu
p\cdot x-\frac 12x\cdot px^\mu -is_0x^\mu -s^{\mu \nu }x_\nu
\end{eqnarray}
where $s^{\mu \nu }=\frac i2(\psi ^\mu \psi ^\nu -\psi ^\nu \psi ^\mu )$.
The operators $x,p,\psi $ are quantum ordered so that the $J^{MN}$ satisfy
the correct algebra for any complex $s_0$. The parameter $s_0$ is an
operator ordering constant which is fixed by hermiticity according to the
Lorenz invariant dot product for states $<\Psi |\Psi >=\int d^{d-1}x\,%
\overline{\Psi }\gamma ^0\Psi $ . Hermiticity $<J^{MN}\Psi |\Psi >=<\Psi
|J^{MN}\Psi >$ fixes $s_0=1/2$. By contrast, in the purely bosonic case we
had $s_0=1$. Thus the presence of the complex $is_0=i/2$ is required for
hermitian $J^{+^{\prime }-^{\prime }},J^{-^{\prime }\mu }$. Furthermore $s_0$
should be consistent with the correct dimension of the Dirac field in $d$
dimensions. When the dimension operator $iJ^{+^{\prime }-^{\prime }}$ is
applied on the Dirac field $<x,\alpha |iJ^{+^{\prime }-^{\prime }}|\Psi
>=\left( x\cdot \partial +\frac 12d-s_0\right) \Psi _\alpha \left( x\right)
\,$ we must obtain $\left( \frac 12d-s_0\right) =\left( d-1\right) /2$ .
Thus we find again $s_0=1/2$. The quadratic Casimir becomes (orbital parts $%
x,p$ drop out) 
\begin{eqnarray}
C_2\left( SO\left( d,2\right) \right) &=&-\frac{d^2}4+s_0^2+\frac 12s^{\mu
\nu }s_{\mu \nu }  \label{C2Lorentz} \\
&=&-\frac{d^2}4+\frac 14+\frac 18\left( d^2-d\right)  \nonumber \\
&=&-\frac 18\left( d+2\right) \left( d-1\right)  \nonumber
\end{eqnarray}
where we have used $\frac 12s^{\mu \nu }s_{\mu \nu }=\frac 12(\psi \cdot
\psi )^2-\frac 14\psi \cdot \psi $ and $\psi \cdot \psi =\gamma \cdot \gamma
/2=d/2$. This agrees precisely with the OSp$\left( 1/2\right) $ gauge
invariance requirements (\ref{c2sod2}) obtained in covariant quantization in
the previous section.

Hence all of the Dirac particle's states correspond to a single and very
special representation of SO$\left( d,2\right) $. This feature is a
reflection of the two-time nature of the spacetime that underlies the Dirac
particle, as is clear in our formulation. As we will see, the same quantum
representation describes many other physical systems by simply choosing
other gauges in the two-time spacetime. In this sense, in the two-time
spacetime, the Dirac particle is dual to all the other physical systems that
we will describe below.

\section{Lightcone gauge \& Harmonic oscillator}

\subsubsection{Free particle in lightcone gauge}

Consider the basis $X^M=\left( X^{+^{\prime }},X^{-^{\prime
}},X^{+},X^{-},X^i\right) $ with the metric $\eta ^{MN}$ taking the values $%
\eta ^{+^{\prime }-^{\prime }}=\eta ^{+-}=-1$ in the lightcone type
dimensions, while $\eta ^{ij}=\delta ^{ij}$ for the remaining $d-2$ space
dimensions. Thus one time $X^{0^{\prime }}$ is a linear combination of $%
X^{\pm ^{\prime }},$ and the other $X^0$ is a linear combination of $X^{\pm
} $. The gauge group OSp$\left(1/2\right) $ has three bosonic and two
fermionic gauge parameters, hence we can make three bosonic and two
fermionic gauge choices. We define the lightcone gauge as $X^{+^{\prime }}=1$%
, $P^{+^{\prime }}=0$, $X^{+}=\tau $, and $\psi ^{+^{\prime }}=\psi ^0=0$.
There is no more gauge freedom left over, so all remaining degrees of
freedom are physical. Inserting this gauge into the constraints (\ref%
{constraints}), and solving them, one finds the following components
expressed in terms of the remaining independent degrees of freedom $%
(x^{-},p^{+},\vec{x}^i\,,\vec{p}^i,\vec{\psi}^i)$%
\begin{eqnarray}
M &=&\left[ +^{\prime }\quad ,\quad \quad -^{\prime }\quad \quad \quad
,\quad \quad +\quad ,\quad -\quad ,\quad i\,\,\,\,\,\right]  \nonumber \\
X^M &=&[1,\,\,\,\,\,\quad (\vec{x}^2/2-\tau x^{-})\,\,\,\,\,\,,\quad \quad
\,\tau \quad \,\,,\quad x^{-},\quad \vec{x}^i\,\,]  \label{relparticl} \\
P^M &=&[0,\,\,(\vec{x}\cdot \vec{p}-x^{-}p^{+}-\frac{\tau \vec{p}^2}{2p^{+}}%
),\,\,\,\,\,p^{+},\,\,\,\,\frac{\vec{p}^2}{2p^{+}}\,\,\,\,,\,\,\,\,\,\,\,%
\vec{p}^i] \\
\psi ^M &=&[0,\,\,\,\,\,\quad \vec{x}\cdot \vec{\psi}-\tau \frac{\,\vec{p}%
\cdot \vec{\psi}}{p^{+}}\,\quad ,\,\,\,\,\,0\quad ,\,\,\,\frac{\,\vec{p}%
\cdot \vec{\psi}}{p^{+}}\,\,\,\,,\,\,\,\,\vec{\psi}^i]
\end{eqnarray}
One can verify that this gauge corresponds to the free relativistic massless
particle, by inserting the gauge fixed form (\ref{relparticl}) into the
action (\ref{actionS}). Since all constraints have been solved, the $%
A^{ij},F^i$ terms drop out, and we get 
\begin{eqnarray}
S_0 &=&\int_0^Td\tau \,\,\left( \partial _\tau X^MP^N\,\eta _{MN}+\frac
i2\psi \cdot \partial _\tau \psi +0+0\right)  \nonumber \\
&=&\int_0^Td\tau \,\,\left( \partial _\tau \vec{x}\cdot \vec{p}-\partial
_\tau x^{-}p^{+}-\frac{\vec{p}^2}{2p^{+}}+\frac i2\psi ^i\partial _\tau \psi
^i\right) .
\end{eqnarray}
This is the action of the free massless spinning relativistic particle in
the lightcone gauge, in the first order formalism, with the correct
Hamiltonian $p^{-}=\vec{p}^2/2p^{+}.$ Note that both time coordinates have
been gauge fixed, $X^{+^{\prime }}=1$ and $X^{+}=\tau $, to describe the
free particle. This is the lightcone ``time''.

The quantization rules are $\left[ x^{-},p^{+}\right] =i\eta ^{+-}=-i$, $%
\left[ \vec{x}^i,\vec{p}^j\right] =i\delta ^{ij}$ and $\left\{ \vec{\psi}^i,%
\vec{\psi}^j\right\} =\delta ^{ij}$. The physical quantum states for $\vec{%
\psi}$ correspond to the basis for the Clifford algebra (with $d-2$
transverse $\vec{\psi}^{\prime }s$). These consist of left spinors of
dimension $2_L^{(d-2)/2-1}$and right spinors of dimension $2_R^{(d-2)/2-1}$
in even dimensions 
\begin{eqnarray}
d &=&12:\quad 16_L\oplus 16_R  \nonumber \\
d &=&10:\quad 8_L\oplus 8_R  \nonumber \\
d &=&8:\quad \,\,\,\,4_L\oplus 4_R  \label{helicities} \\
d &=&6:\quad \,\,\,\,2_L\oplus 2_R\quad  \nonumber \\
d &=&4:\quad \,\,\,\,1_L\oplus 1_R  \nonumber
\end{eqnarray}
For odd dimensions one gets the sum of the $L$ and $R$ spinors of the lower
even dimension. These are the helicity states for massless fermions from the
lightcone point of view. For example in four dimensions there is one degree
of freedom for a massless left handed ``neutrino'' and one degree of freedom
for a massless right handed ``neutrino''.

The SO$\left( d,2\right) $ generators of eq.(\ref{confgen}) now take the
form (at $\tau =0$) 
\begin{eqnarray}
&&\left. J^{ij}=\vec{x}^i\vec{p}^j-\vec{x}^j\vec{p}^i+S^{ij}\right. \quad
\quad \quad \quad \quad \quad \quad \quad \quad \quad SO\left( d-2\right) \\
&&\left. 
\begin{array}{l}
J^{+^{\prime }-}=\frac{\vec{p}^2}{2p^{+}},\quad J^{+-}=-\frac 12\left(
x^{-}p^{+}+p^{+}x^{-}\right) , \\ 
J^{-^{\prime }+}=\frac 12\vec{x}^2p^{+},\quad J^{+^{\prime }+}=p^{+} \\ 
J^{+^{\prime }-^{\prime }}=\frac 12\left( \vec{x}\cdot \vec{p}\mathbf{+}\vec{%
p}\cdot \vec{x}-x^{-}p^{+}-p^{+}x^{-}\right) \\ 
J^{-^{\prime }-}=\left[ 
\begin{array}{l}
\frac 1{8p^{+}}\left( \vec{x}^2\vec{p}^2+\vec{p}^2\vec{x}^2-2\alpha \right)
\\ 
-\frac{x^{-}}2\left( \vec{x}\cdot \vec{p}\mathbf{+}\vec{p}\cdot \vec{x}%
\right) +x^{-}p^{+}x^{-} \\ 
+\frac 1{2p^{+}}\left( \vec{x}^i\vec{p}^j-\vec{x}^j\vec{p}^i\right) S^{ij}%
\end{array}
\right]%
\end{array}
\right\} \quad SO\left( 2,2\right)  \label{freewithspin} \\
&&\left. J^{+^{\prime }i}=\vec{p}^i,\quad \quad \quad \quad J^{+i}=-\vec{x}%
^ip^{+},\quad \right. \\
&&\left. J^{-i}=x^{-}\vec{p}^i-\frac 1{2p^{+}}\vec{p}^j\vec{x}^i\vec{p}%
^j-\frac 1{p^{+}}S^{ij}\vec{p}_j\right. \\
&&\left. J^{-^{\prime }i}=\left[ 
\begin{array}{l}
\frac 12\vec{x}^j\vec{p}^i\vec{x}^j-\frac 12\vec{x}\cdot \vec{p}\vec{x}%
^i-S^{ij}\vec{x}^j \\ 
-\frac 12\vec{x}^i\vec{p}\cdot \vec{x}+\frac 12\vec{x}^i\left(
x^{-}p^{+}+p^{+}x^{-}\right)%
\end{array}
\right] \right.
\end{eqnarray}
where 
\begin{equation}
S^{ij}=\frac 1{2i}\left[ \vec{\psi}^i,\vec{\psi}^j\right]
\end{equation}
is a spin operator in the transverse dimensions. The quantum operators are
ordered so that all generators $J^{MN}$ are hermitian. The constant $\alpha $
that appears in $J^{-^{\prime }-}$ arises due to quantum operator ordering
ambiguities. It is fixed to $\alpha =2-d$ by demanding the correct closure
for the commutator 
\begin{equation}
\left[ L^{-^{\prime }i},L^{-\,j}\right] =i\delta ^{ij}L^{-^{\prime }-},\quad
\rightarrow \quad \alpha =2-d.
\end{equation}
By contrast, in the purely bosonic case we had a similar $\alpha _{bose}=-1$ 
\cite{dualconf, dualH} .

The Hilbert space may be labelled by the commuting momentum operators of the
free particle as well as its spin in the form of helicity states as given
above in (\ref{helicities}) 
\begin{equation}
|\,\vec{p},p^{+},p^{-}=\vec{p}^2/2p^{+};helicities>.
\end{equation}
This is the free particle Hilbert space, which is complete. It is unitary
and has the usual delta function normalization. Wave packets with finite
positive norm are constructed as usual, and they correspond to the solutions
of the Dirac equation of the previous section written in lightcone
coordinates. The operators $J^{MN}$ given above act on the states in a
natural way and these states form a basis for SO$\left( d,2\right) .$ The
Casimir eigenvalues are easily computed directly by squaring the operators.
By using our previous calculation of the purely bosonic case \cite{dualconf,
dualH}, and the property $\vec{\psi}\cdot \vec{\psi}=\left( d-2\right) /2$,
we find (see also below) 
\begin{equation}
\frac 12J^{MN}J_{MN}=-\frac 18\left( d+2\right) \left( d-1\right) ,
\end{equation}
in agreement with fully covariant quantization and Lorentz covariant
quantizations given in the previous sections.

The interpretation of the physics in this basis of SO$\left( d,2\right) $
is, of course, the same as the previous section. Next we show that the same
construction of SO$\left( d,2\right) $ has a different physical
interpretation.

\subsection{Harmonic oscillator with spin}

The same realization of SO$\left( d,2\right) $, with the same eigenvalues of
the Casimirs, is also related to the Harmonic oscillator. In (\ref%
{freewithspin}) the SO$\left( 2,2\right) $ subgroup which is equivalent to $%
SL\left( 2,R\right) _L\otimes SL\left( 2,R\right) _R$ has the following
generators $G_{0,1,2}^{L,R}$ with the standard algebra $\left[ G_0,G_1\right]
=iG_2$, $\left[ G_0,G_2\right] =-iG_1$, $\left[ G_1,G_2\right] =-iG_0$, 
\begin{eqnarray}
SL\left( 2,R\right) _R &:&G_2^R=\frac 12\left( J^{+^{\prime }-^{\prime
}}-J^{+-}\right) ,\quad G_0^R\pm G_1^R=J^{\pm ^{\prime }\mp }.
\label{sl2sl2} \\
SL\left( 2,R\right) _L &:&G_2^L=\frac 12\left( J^{+^{\prime }-^{\prime
}}+J^{+-}\right) ,\quad G_0^L\pm G_1^L=J^{\pm ^{\prime }\pm }\,\,,
\end{eqnarray}
Thus the compact generator $G_0^R$ of SL$\left( 2,R\right) _R$ is given by
the harmonic oscillator Hamiltonian 
\begin{equation}
G_0^R=\frac 12\left( J^{+^{\prime }-}+J^{-^{\prime }+}\right) =\frac{\vec{p}%
^2}{4p^{+}}+\frac 14\vec{x}^2p^{+}.
\end{equation}
The mass of the harmonic oscillator is $M=2p^{+}$ and the frequency is $%
\omega =1/2$. The mass is given by the generator $J^{+^{\prime
}+}=p^{+}=G_0^L+G_1^L$ of SL$\left( 2,R\right) ^L$.

Even though the particle has spin degrees of freedom, the harmonic
oscillator Hamiltonian is independent of spin. For a fixed mass $M=2p^{+}$,
its quantum eigenstates $|p^{+},E_n,l,s,j>$ are labelled with the
eigenvalues of energy $E=G_0^R$, orbital and spin angular momentum $l,s$
and/or $j$ for total $SO\left( d-2\right) $ spin $J^{ij}$. Of course, from
the solution of the harmonic oscillator quantum mechanics in $d-2$ space
dimensions, we already know that the energy quantum numbers should be $%
E=\omega (n+\frac 12(d-2))=\frac n2+\frac 14(d-2)$, with the angular
momentum also determined 
\begin{eqnarray}
n &=&0,1,2,\cdots \\
l &=&n,(n-2),(n-4),\cdots ,(0\,\,or\,1).
\end{eqnarray}
The degeneracy of the state at level $n$ corresponds to an SU$\left(
d-2\right) $ multiplet described by a single row Young tableau with $n$
boxes, times the degeneracy of the spin states which is the same at every
level. This Young tableau decomposed under SO$\left( d-2\right) $ gives
completely symmetric traceless tensors with $l$ indices $T_{i_1i_2\cdots
i_l}\left( \vec{x}\right) $, with the values of $l$ indicated above. The
total $J^{ij}$ SO$\left( d-2\right) $ spin $j$ is obtained by combining the
orbital and spin parts for SO$\left( d-2\right) $. To make a connection to
the group theory below it is useful to rewrite $n=l+2n_r$ where both $l,n_r$
are positive integers, and $n_r$ has the meaning of radial quantum number.
So we may write the energy eigenvalue in the form 
\begin{equation}
E=G_0^R=\frac 14(d-2)+\frac l2+n_r{.}  \label{harmoE}
\end{equation}

Now we explain how these harmonic oscillator quantum numbers fully label the
same unique representation of SO$\left( d,2\right) $, and how the full set
of harmonic oscillator states at all energy levels provide a \textit{single}
irreducible representation. The key here is that the mass 2$p^{+}$ as well
as the spin are labels of the representation and they must transform under SO%
$\left( d,2\right) $. In this sense the mass is the analog of a modulus
parameter that transforms under duality. Furthermore, the choice of $G_0^R$
as Hamiltonian implies a different choice of time as embedded in $d+2$
dimensions, as compared to the free particle time.

A basis for the group theory representation space is labelled by the SO$%
\left( d,2\right) $ Casimir eigenvalues, and the $SO\left( d-2\right)
\otimes \,SL\left( 2,R\right) _L\otimes \,SL\left( 2,R\right) _R$ subgroups 
\begin{eqnarray}
|Casimirs;\,SO\left( d-2\right) ;\,SL\left( 2,R\right) _L;\,SL\left(
2,R\right) _R &>&=  \label{states} \\
|Casimirs;\,l\,,s;\,j_Lp^{+};j_Rm_R &>&  \nonumber
\end{eqnarray}
The SL$\left( 2,R\right) _L\otimes \,SL\left( 2,R\right) _R$ subspace is
labelled by $|j_Lp^{+};j_Rm>,$ where $m$ is the eigenvalue of the compact
generator of SL$\left( 2,R\right) _R$ that coincides with the Hamiltonian $%
G_0^R=m=E,$ and $p^{+}$ is the eigenvalue of the SL$\left( 2,R\right) _L$
generator $J^{+^{\prime }+}=G_0^L+G_1^L=p^{+}$. We will compare these
quantum numbers to those of the harmonic oscillator given above.

First we compare $m$ to the energy eigenvalue $E$. The quantum number $m$ is
determined from representation theory of SL$\left( 2,R\right) $. Since $%
G_0^R $ is a positive operator, the only possible representation is the
positive discrete series, for which $E=G_0^R=m=j_R+1+n_r$ with $%
n_r=0,1,2,\cdots $. There remains to show that $j_R+1$ is the remaining part
of eq.(\ref{harmoE}), which we will do below.

The $SO\left( d-2\right) $ quantum numbers $\left( l,s\right) $ are
determined by orbital $l$ and spin quantum numbers $s$. In the construction
of SO$\left( d,2\right) $ given in (\ref{freewithspin}) orbital angular
momentum $L^{ij}$ can only have representations labelled by integers $l$
that corresponds to the completely symmetric traceless tensor $%
T_{i_1i_2\cdots i_l}\left( \vec{x}\right) $ with $l$ indices in $\left(
d-2\right) $ dimensions. Similarly $s$ is limited to the spinor
representations listed in (\ref{helicities}). The direct product of these
representations is what is symbolized by the quantum numbers $\left(
l,s\right) $. So, these are the same angular momentum labels as the harmonic
oscillator.

There remains to specify the values of $j_L,j_R$. They are computed through
the Casimirs 
\begin{equation}
j_{L,R}(j_{L,R}+1)=\left( G_0^{L,R}\right) ^2-\left( G_1^{L,R}\right)
^2-\left( G_2^{L,R}\right) ^2.
\end{equation}
Using the $\vec{x},\vec{p},\vec{\psi}$ representation for $G_{0,1,2}^{L,R}$
given in (\ref{freewithspin},\ref{sl2sl2}) we find that they $j_{L,R}$ are
not independent of the orbital and spin angular momenta 
\begin{eqnarray}
j_R(j_R+1) &=&\frac 18L_{ij}L^{ij}+\frac 1{16}\left( d-2\right) \left(
d-6\right) \\
j_L(j_L+1) &=&\frac 18L_{ij}L^{ij}+\frac 1{16}\left( d-2\right) \left(
d-6\right) +\frac 12L_{ij}S^{ij} \\
&=&\frac 14J_{ij}J^{ij}-\frac 18L_{ij}L^{ij}-\frac 3{16}\left( d-2\right) . 
\nonumber
\end{eqnarray}
The allowed eigenvalues for SO$\left( d-2\right) $ orbital angular momentum
are $\frac 12L_{ij}L^{ij}=l\left( l+d-4\right) $ (completely symmetric
traceless tensor with $l$ indices in $d-2$ dimensions), and the allowed
values of SO$\left( d-2\right) $ spin are $\frac 12S_{ij}S^{ij}=\frac
18\left( d-2\right) \left( d-3\right) $ (from $\vec{\psi}\cdot \vec{\psi}%
=\left( d-2\right) /2$). From these we deduce the allowed values of $j_R,j_L$
and SO$\left( d-2\right) $ total angular momentum $j$, 
\begin{equation}
j_R(j_R+1)=\frac 14l\left( l+d-4\right) +\frac 1{16}\left( d-2\right) \left(
d-6\right)
\end{equation}
gives 
\begin{equation}
j_R=\frac 12l+\frac 14d-\frac 32,\quad l=0,1,2,\cdots
\end{equation}
Similarly, we obtain $j_L$ for $d=5$ 
\begin{equation}
j_{L_{d=5}}=\left\{ 
\begin{array}{c}
d=5:SO\left( 3\right) \quad s=\pm 1/2 \\ 
\frac 12J_{ij}J^{ij}=j\left( j+1\right) \\ 
\,j=l\pm 1/2,\,\,\,\,j=\frac 12^{+},\frac 32^{+},\frac 32^{-},\frac
52^{+},\frac 52^{-},\cdots \, \\ 
j_L(j_L+1)=\frac 12\left( l+s\right) \left( l+s+1\right) -\frac 14l\left(
l+1\right) -\frac 9{16} \\ 
\,j_L=-\frac 12+\frac 12\sqrt{\left( j+\frac 12\right) \left( j+\frac 12\pm
1\right) -2}\,%
\end{array}
\quad \right.
\end{equation}
and $d=6$%
\begin{equation}
j_{L_{d=6}}=\left\{ 
\begin{array}{c}
d=6:SO\left( 4\right) =SU\left( 2\right) _L\otimes SU\left( 2\right) _R\quad
s_{L,R}=\pm 1/2 \\ 
\frac 12J_{ij}J^{ij}=2j_1\left( j_1+1\right) +2j_2\left( j_2+1\right) \\ 
\left( j_1,j_2\right) =\left( \frac l2\pm \frac 12,\frac l2\right) \oplus
\left( \frac l2,\frac l2\pm \frac 12\right) \\ 
=\left( j,j\mp \frac 12\right) \oplus \left( j\mp \frac 12,j\right) \quad
j=\frac 12,1,\frac 32,2,\cdots \\ 
j_L(j_L+1)=\left( \frac l2+s\right) \left( \frac l2+s+1\right) -\frac 34 \\ 
j_L(j_L+1)=j\left( j+1\right) -\frac 34 \\ 
j_L=-\frac 12+\frac 12\sqrt{\left( 2j+1\right) ^2-3}\,%
\end{array}
\quad \right.
\end{equation}
For other values of $d$ the computation of $j_L$ is a technical matter. This
verifies that the energy eigenvalue and other quantum numbers coincide with
the group theoretical representation labels. The only independent labels are
those of the harmonic oscillators, including mass and spin, while the
remaining group theory labels are determined by them. Among the group theory
labels we must include the mass $M=2p^{+}$.

The realization of SO$\left( d,2\right) $ on this harmonic oscillator system
is quite non-trivial. As already verified, the Casimir eigenvalues for SO$%
\left( d,2\right) $ are the ones determined by OSp$\left( 1/2\right) $ gauge
invariance in (\ref{c2sod2}). The choice of time as embedded in $d+2$
dimensions has a different topology than the free particle. The quantum
space is dual to the free particle, while both systems represent the same
two-time quantum theory in unitarily equivalent bases.

\section{``H-atom'' with spin}

The free Dirac particle may be described in the $x^0=\tau $ gauge (see next
section) instead of the $x^{+}=\tau $ gauge of the previous section. To
describe the H-atom we take a gauge that is dual to the free Dirac particle.
The duality relation to the free particle is obtained by flipping the roles
of $\mathbf{r,p}$ followed by a discrete Sp$\left( 2\right) $
transformation. The resulting gauge is (at fixed time $\tau =0$) $%
X^{+^{\prime }}=0,\,P^{+^{\prime }}=1,\,P^0=0,\psi ^{+^{\prime }}=0\,$ 
\begin{eqnarray}
M &=&\left( +^{\prime },\mathbf{-}^{\prime },0,i\right)  \nonumber \\
X^M &=&\left( 0,\mathbf{r\cdot p},r,\mathbf{r}^i\right)  \label{Hsecond} \\
P^M &=&\left( 1,\frac{\mathbf{p}^2}2,0,\mathbf{p}^i\right) \\
\psi &=&\left( 0,\mathbf{\psi \cdot p},\frac 1r\mathbf{\psi \cdot r},\mathbf{%
\psi }^i\right) .
\end{eqnarray}
where $i=1,2,\cdots ,\left( d-1\right) $. All constraints, $X^2=P^2=X\cdot
P=X\cdot \psi =\psi \cdot P=0,$ are explicitly solved. The generators of the
conformal group (\ref{confgen}) take the form (recall $\eta ^{+^{\prime
}-^{\prime }}=\eta ^{00}=-1$) 
\begin{eqnarray}
J^{-^{\prime }+^{\prime }} &=&\frac 12\left( \mathbf{r\cdot p+p\cdot r}%
\right) \\
J^{0+^{\prime }} &=&r,\quad J^{i+^{\prime }}=\mathbf{r}^i \\
J^{0-^{\prime }} &=&\frac 12\mathbf{p}^ir\mathbf{p}^i+\frac ar+\frac 1{2r}%
\mathbf{S}_{ij}\mathbf{L}^{ij} \\
J^{i-^{\prime }} &=&-\frac 12\mathbf{p\cdot \,r\,\,p}^i-\frac 12\mathbf{p}^i%
\mathbf{\,r\cdot p\,\,}+\frac 12\mathbf{p}^j\mathbf{r}^i\mathbf{p}^j+b\frac{%
\mathbf{r}^i}{r^2}+\mathbf{S}^{ij}\mathbf{p}_j \\
J^{i0} &=&-\frac 12\left( r\mathbf{p}^i+\mathbf{p}^ir\right) +\frac 1r%
\mathbf{S}^{ij}\mathbf{r}_j \\
J^{ij} &=&\mathbf{r}^i\mathbf{p}^j-\mathbf{r}^j\mathbf{p}^i+\mathbf{S}^{ij}
\end{eqnarray}
where 
\begin{equation}
\mathbf{S}^{ij}=\frac 1{2i}\left( \psi ^i\psi ^j-\psi ^j\psi ^i\right) .
\end{equation}
As in the purely bosonic case \cite{dualH} there are ordering ambiguities
represented by the constants $a,b$ that appear in $J_{0+^{\prime }}$ and $%
J_{+^{\prime }i}$ respectively. By using the basic commutation relations
among $(\mathbf{r,p,\psi )}$ one can check that the SO$\left( d,2\right) $
commutation relations are indeed satisfied for any $a,$ while $b$ is fixed
to $b=-a$ by demanding correct closure for the commutator 
\begin{equation}
\left[ J^{0-^{\prime }},J^{i0}\right] =-iJ^{i-^{\prime }}\quad \rightarrow
b=-a
\end{equation}
By contrast, in the purely bosonic case we had $b_{bose}=-a_{bose}-\frac{d-2}%
4$. The remaining parameter $a$ will be fixed by the OSp$\left( 1/2\right) $
gauge invariance, not by the SO$\left( d,2\right) $ algebra, as will be
discussed below.

It is evident that the operators $\mathbf{J}^{ij}$ form the algebra of the
rotation subgroup SO$\left( d-1\right) $. Its quadratic Casimir is given by 
\begin{eqnarray}
\frac 12\mathbf{J}_{ij}\mathbf{J}^{ij} &=&\frac 12\mathbf{L}_{ij}\mathbf{L}%
^{ij}+\mathbf{L}_{ij}\mathbf{S}^{ij}+\frac 12\mathbf{S}_{ij}\mathbf{S}^{ij} 
\nonumber \\
&\equiv &\left( \mathbf{r}^j\mathbf{p}^2\mathbf{r}^j-\mathbf{r\cdot
p\,p\cdot \,r}\right) \mathbf{+L}_{ij}\mathbf{S}^{ij} \\
&&+\frac 14\left( 2\left( \mathbf{\psi }\cdot \mathbf{\psi }\right) ^2-%
\mathbf{\psi \cdot \psi }\right)  \nonumber
\end{eqnarray}
Similarly, the following three operators form a SO$\left( 1,2\right) $
subalgebra 
\begin{eqnarray}
J^{-^{\prime }+^{\prime }} &\equiv &J_2,\quad J^{0-^{\prime }}\equiv \frac
12\left( J_0+J_1\right) ,\quad J^{0+^{\prime }}\equiv J_0-J_1, \\
J_2 &=&\frac 12\left( \mathbf{r\cdot p+p\cdot r}\right) ,\quad \left(
J_0+J_1\right) =\mathbf{p}^ir\mathbf{p}^i+\frac{2a}r+\frac 1r\mathbf{S}_{ij}%
\mathbf{L}^{ij},\quad J_0-J_1=r  \nonumber
\end{eqnarray}
For any $a$ they close correctly 
\begin{equation}
\left[ J_0,J_1\right] =iJ_2,\quad \left[ J_0,J_2\right] =-iJ_1,\quad \left[
J_1,J_2\right] =-iJ_0.
\end{equation}
The compact generator $J_0$ is given in terms of the canonical operators as 
\begin{equation}
J_0=J^{0-^{\prime }}+\frac 12J^{0+^{\prime }}=\frac 12\mathbf{p}^ir\mathbf{p}%
^i+\frac ar+\frac r2+\frac 1{2r}\mathbf{S}_{ij}\mathbf{L}^{ij}.
\label{compact}
\end{equation}
The quadratic Casimir operator for this subalgebra takes the form 
\begin{eqnarray}
J(J+1) &=&J_0^2-J_1^2-J_2^2=J^{0-^{\prime }}J^{0+^{\prime }}+J^{0+^{\prime
}}J^{0-^{\prime }}-\left( J^{-^{\prime }+^{\prime }}\right) ^2  \nonumber \\
&=&\frac 12\mathbf{L}_{ij}\mathbf{L}^{ij}+\mathbf{S}_{ij}\mathbf{L}%
^{ij}+\frac 14\left( d-2\right) ^2-\frac 14+2a  \nonumber \\
&=&\frac 12\mathbf{J}_{ij}\mathbf{J}^{ij}-\frac 12\mathbf{S}^{ij}\mathbf{S}%
_{ij}+\frac 14\left( d-2\right) ^2-\frac 14+2a  \nonumber \\
&=&\frac 12\mathbf{J}_{ij}\mathbf{J}^{ij}+\frac 18\left( d-1\right) \left(
d-4\right) +2a  \label{JJ}
\end{eqnarray}
where we have used $\frac 12\mathbf{S}^{ij}\mathbf{S}_{ij}=\frac 12(\mathbf{%
\psi }\cdot \mathbf{\psi )}^2-\frac 14\mathbf{\psi \cdot \psi }$ and $%
\mathbf{\psi \cdot \psi =}(d-1)/2$. We see that the quadratic Casimir
operators of the SO$\left( 1,2\right) $ subalgebra and that of the rotation
subgroup SO$\left( d-1\right) $ are related to each other in this
representation of SO$\left( d,2\right) $. The overall quadratic Casimir
operator for SO$\left( d,2\right) $ may now be evaluated. All orbital parts $%
\mathbf{r,p}$ drop out, and the result is 
\begin{eqnarray}
C_2 &=&\frac 12J_{MN}J^{MN}  \nonumber \\
&=&-\left( J^{-^{\prime }+^{\prime }}\right) ^2+J^{0-^{\prime
}}J^{0+^{\prime }}+J^{0+^{\prime }}J^{0-^{\prime }}  \nonumber \\
&&-J^{i-^{\prime }}J^{i+^{\prime }}-J^{i+^{\prime }}J^{i-^{\prime
}}-J^{i0}J^{i0}+\frac 12\mathbf{J}_{ij}\mathbf{J}^{ij}  \label{Hso12c2} \\
&=&-\frac 18d^2-\frac 18d-\frac 34+4a  \nonumber \\
&=&-\frac 18\left( d+2\right) \left( d-1\right) \quad \rightarrow a=\frac 14
\nonumber
\end{eqnarray}
We see that $a=\frac 14$ is fixed by the requirement of OSp$\left(
1/2\right) $ gauge invariance (\ref{c2sod2}) that was obtained in covariant
quantization. Therefore the last step fixes the values of $a$ and $b$
uniquely in the gauge invariant sector 
\begin{equation}
a=-b=\frac 14\,.
\end{equation}
These values correspond to the following quantum ordering of the operators
in $J^{0-^{\prime }}$ 
\begin{eqnarray}
J^{0-^{\prime }} &=&\frac 12\mathbf{p}^ir\mathbf{p}^i+\frac 1{4r}+\frac 1{2r}%
\mathbf{S}_{ij}\mathbf{L}^{ij} \\
&=&r^{\frac 12}\left[ \frac 12\mathbf{p}^2\right] r^{\frac 12}-\frac
1{8r}\left( 3-2d\right) +\frac 1{2r}\mathbf{S}_{ij}\mathbf{L}^{ij}. 
\nonumber
\end{eqnarray}

We now proceed to solve the system algebraically, and show its relation to
the H-atom. A basis for the quantum theory is chosen to diagonalize the
Hamiltonian. In our case we will show that this corresponds to the SO$\left(
d,2\right) $ representation basis labelled by the subgroups 
\begin{equation}
|Casimirs;SO\left( d-1\right) ;SO\left( 1,2\right) >,
\end{equation}
and that all the states of the ``H-atom'' with spin correspond to a single
irreducible representation of SO$\left( d,2\right) $, with the Casimirs
given before by the covariant quantization (i.e. OSp$\left( 1/2\right) $
gauge invariance).

As explained earlier, since we have two timelike dimensions, the choice of
``time'' corresponds to a choice of Hamiltonian as a combination of the
generators of SO$\left( d,2\right) $. One such choice is dual to another via
OSp$\left( 1/2\right) $ gauge transformations. We now make the following
choice for ``Hamiltonian'' $h=J^{0^{\prime }0}=J_0$ which is the compact
generator of the SO$\left( 1,2\right) $ subgroup. One way of justifying this
gauge choice is the algebraic demonstration below that it corresponds to the 
$1/r$ potential. Another way is to choose a (canonically related) gauge in
which the original action reduces to the interacting system with $1/r$
potential, and then show that $J^{0^{\prime }0}$ is related to the
Hamiltonian. This was done explicitly in \cite{dualH} for the spinless case.
See also the last paragraph of section (6.2). Thus, consider $h=J^{0^{\prime
}0}=J_0$ in the form 
\begin{eqnarray}
h &=&J_0=J^{0-^{\prime }}+\frac 12J^{0+^{\prime }}  \nonumber \\
&=&r^{\frac 12}\left[ \frac 12\mathbf{p}^2+\frac 12-\frac{\left( 3-2d\right) 
}{8r^2}+\frac 1{2r^2}\mathbf{S}_{ij}\mathbf{L}^{ij}\right] r^{\frac 12} 
\nonumber \\
&=&r^{\frac 12}\left[ 
\begin{array}{c}
\frac 12\left( p_r^2+\frac 1{2r^2}\mathbf{L}^{ij}\mathbf{L}_{ij}+\frac
1{4r^2}\left( d-2\right) \left( d-4\right) \right) \\ 
+\frac 12-\frac{\left( 3-2d\right) }{8r^2}+\frac 1{2r^2}\mathbf{S}_{ij}%
\mathbf{L}^{ij}%
\end{array}
\right] r^{\frac 12} \\
&=&r^{\frac 12}\left[ \frac 12\left( p_r^2+\frac 1{r^2}\left( \frac 12%
\mathbf{J}^{ij}\mathbf{J}_{ij}-\frac 12\mathbf{S}^{ij}\mathbf{S}_{ij}+\frac
14d^2-\frac 44d+\frac 54\right) \right) +\frac 12\right] r^{\frac 12} 
\nonumber \\
&=&r^{\frac 12}\left[ \frac 12\left( p_r^2+\frac 1{r^2}\left( \frac 12%
\mathbf{J}^{ij}\mathbf{J}_{ij}+\frac 18\left( d^2-5d+8\right) \right)
\right) +\frac 12\right] r^{\frac 12}  \nonumber \\
&=&r^{\frac 12}\left[ \frac 12\left( p_r^2+\frac 1{r^2}J(J+1)\right) +\frac
12\right] r^{\frac 12}\,\,\,\,.  \nonumber
\end{eqnarray}
Here we have used $\mathbf{p}^2=p_r^2+\frac 1{2r^2}\mathbf{L}^{ij}\mathbf{L}%
_{ij}+\frac 1{4r^2}\left( d-2\right) \left( d-4\right) $ where $p_r=\frac
1{2r}\mathbf{r\cdot p+p\cdot r}\frac 1{2r}$ is the hermitian radial momentum
canonically conjugate to $r$, and have shown that $J(J+1)$, which is the
quadratic Casimir of $SO(1,2)$ as given in (\ref{JJ}), emerges in the radial
equation. From here one may proceed in two ways. Either one may solve the
radial equation given below directly, or use an algebraic approach. The
agreement between the two is a check of our calculation.

We proceed with the algebraic approach. Since $h=J_0$ is a generator of the
SO$\left( 1,2\right) $ algebra it is diagonalized on the usual SO$\left(
1,2\right) $ basis $|Jm>$ where $m$ is the quantized eigenvalue of the
compact generator $J_0$. Evidently the operator $h$ is positive, therefore $%
m $ can only be positive. This is possible only in the positive unitary
discrete series representation of SO$\left( 1,2\right) ,$ and therefore the
spectrum of $m$ must be 
\begin{equation}
m=J+1+n_r,\quad n_r=0,1,2,\cdots .
\end{equation}
where, as we will see shortly, the integer $n_r$ will play the role of the
radial quantum number.

Let us now show the relation to the Hydrogen atom Hamiltonian. Applying $h$
on these states we have 
\begin{equation}
r^{\frac 12}\left[ \frac 12\mathbf{p}^2+\frac 12+.....\right] r^{\frac
12}|Jm>=m|Jm>
\end{equation}
Multiplying it with the operator $r^{-\frac 12}$ from the left, this
equation is rewritten as 
\begin{equation}
\left[ \frac 12\mathbf{p}^2+\frac 12-\frac mr+.....\right] \left( r^{\frac
12}|Jm>\right) =0.
\end{equation}
We now recognize that the states $|\psi _m>=\left( r^{\frac 12}|jm>\right) $
are eigenstates of the Hydrogen atom Hamiltonian. Actually this is a
rescaled form of the standard Hamiltonian equation written in terms of
dimensionful coordinates and momenta $\mathbf{\tilde{r},\tilde{p}}$%
\begin{equation}
\left[ \frac{\mathbf{\tilde{p}}^2}{2M}-\frac \alpha {\tilde{r}}+....\right]
|\psi _m>=E_m|\psi _m>.
\end{equation}
The quantized energy is 
\begin{equation}
E_n=-\frac{\alpha ^2}{m^2}=-\frac{\alpha ^2}{\left( J+1+n_r\right) ^2}.
\end{equation}
We still need to figure out the values of $J$ as a function of the quantized
integers $l$ and $n_r$. As an example consider $d=4$, SO$\left( d-1\right)
=SO\left( 3\right) $, for which we are familiar with the use of addition of
angular momentum. Adding spin 1/2 to orbital quantum number $l$, gives $%
j=l\pm 1/2$ with $l=0,1,2,\cdots $. From these we compute $J\left( j\right) $
by inserting $j(j+1)=\frac 12J_{ij}J^{ij}$ in (\ref{JJ}) 
\begin{eqnarray}
J\left( j\right) &=&-\frac 12+\frac 12\sqrt{\left( 3+4j^2+4j\right) }\quad
j=\frac 12,\frac 12,\frac 32,\frac 32,\frac 52,\frac 52,\cdots  \nonumber \\
&=&-\frac 12+\frac 12\sqrt{\left( 6+4l^2+8l\right)
_{j=l+1/2}\,\,\,or\,\,\,\left( 2+4l^2\right) _{j=l-1/2}}
\end{eqnarray}
The energy depends on both on $j$ and $n_r$ and therefore there is no
accidental degeneracies, like the SO$\left( 4\right) $ in 4D$.$ For other
dimensions $d$ we compute $J$ by using a similar procedure.

Although we have named this gauge the ``H-atom'' gauge (with quotes !)
because of the $1/r$ potential, evidently the system does not describe the
usual H-atom with the usual spin correction, since the spin dependence has a
different $r$ dependence than the usual $\mathbf{L\cdot S}$ correction (here 
$1/r^2$, whereas the usual correction is $1/r^3$). Nevertheless, as seen
below, it is possible to find a gauge that gives any interacting
non-relativistic Hamiltonian, including the correct spin correction for the
H-atom. However, the SO$\left( d,2\right) $ representation becomes
considerably more complicated and the quantum ordering issues for all the
generators become technically difficult to resolve. Our aim here was to show
that for the cases for which we could resolve the quantum ordering, the
model does correspond to the same representation of SO$\left( d,2\right) $
in all gauges, and hence the corresponding physical systems are dual in this
sense in the quantum theory.

\section{More interactions as gauge choices}

The action in a general gauge is given in (\ref{actionS}). We will show that
we can construct more general interacting non-relativistic system in $d-1$
space dimensions, with Hamiltonian of the form 
\begin{equation}
H=\frac{\mathbf{p}^{2}}{2}+V,
\end{equation}%
with more general potential function $V$ (within a class noted below) and SO$%
\left( d-1\right) $ spin $\frac{1}{2}S_{ij}S_{ij}=\frac{1}{8}\left(
d-1\right) \left( d-2\right) $, simply by taking appropriate gauge choices
for time. We must emphasize that we expect that there are more gauge choices
that would yield other forms of Hamiltonians.

\subsection{Free spinning particle in timelike gauge}

First let us remind ourselves of the method by reconsidering the free
massless particle of (\ref{relativistic}) in the timelike gauge ($x^0\left(
\tau \right) =\tau $ ). The same method will be applied to the more general
case. The following parametrization solves all of the contraints (\ref%
{constraints}) 
\begin{eqnarray}
M &=&\left( +^{\prime }\quad ,\quad \mathbf{-}^{\prime }\quad \quad ,\quad
0\quad ,\quad i\right)  \nonumber \\
X^M &=&\left( 1,\,\,\,\frac 12\left( \mathbf{r}^2-\tau ^2\right) ,\,\quad
\tau \,\quad ,\,\,\mathbf{r}^i\right) \\
P^M &=&\left( 0,\,\,\,\mathbf{r\cdot p-\left| p\right| }\tau ,\quad \left| 
\mathbf{p}\right| \quad ,\,\,\mathbf{p}^i\right)  \label{timelike} \\
\psi &=&\left( 0,\,(\mathbf{r\cdot \,\psi -}\frac{\mathbf{r\cdot p}}{^{%
\mathbf{p}^2}}\mathbf{\,p\cdot \psi )\,\,\,},\,0,\mathbf{\psi }^i-\frac{%
\mathbf{p}^i}{^{\mathbf{p}^2}}\mathbf{p\cdot \psi }\right) \\
&&+\chi X^M+\xi P^M  \nonumber
\end{eqnarray}
where $\chi ,\xi $ represents fermionic gauge freedom, and they can be
chosen at convenience so as to obtain the simplest possible action or SO$%
\left( d,2\right) $ generators. This gauge is OSp$\left( 1/2\right) $ dual
to the H-atom gauge (\ref{Hsecond}) used in the previous section. At $\tau
=0 $, the duality transformation consists of choosing $\chi =0$ and $\xi
=-\frac 1{^{\mathbf{p}^2}}\mathbf{p\cdot \psi },$ plus a discrete Sp$\left(
2\right) $ transformation that interchages $X^M,P^M,$ and then re-naming $%
\mathbf{r\leftrightarrow p}$. By inserting this gauge choice, with $\xi
=\chi =0 $, into the action (\ref{actionS}) we obtain 
\begin{eqnarray}
S_0 &=&\int_0^Td\tau \,\,\left[ P\cdot \partial _\tau X+\frac i2\psi \cdot
\partial _\tau \psi +0+0\right]  \nonumber \\
&=&\int_0^Td\tau \,\,\left[ \mathbf{\dot{r}\cdot p}-\left| \mathbf{p}\right|
+\frac i2\mathbf{\tilde{\psi}}^i\partial _\tau \mathbf{\tilde{\psi}}^j\right]
\,,
\end{eqnarray}
where 
\begin{equation}
\mathbf{\tilde{\psi}}^i=\left( \delta _{ij}-\frac{\mathbf{p}_i\mathbf{p}_j}{%
\mathbf{p}^2}\right) \mathbf{\psi }^j\,.
\end{equation}
This action describes the free massless relativistic spinning particle, with
Hamiltonian $H=\left| \mathbf{p}\right| $. Using Noether's theorem, we see
that the generator of rotations is $J^{ij}=L^{ij}+\tilde{S}^{ij}$, where $%
\tilde{S}^{ij}=\frac 1{2i}\left( \mathbf{\tilde{\psi}}^i\mathbf{\tilde{\psi}}%
^j-\mathbf{\tilde{\psi}}^j\mathbf{\tilde{\psi}}^i\right) $. Only the spin
components perpandicular to momentum can appear since $\mathbf{\tilde{\psi}}%
\cdot \mathbf{p}=0 $. This is as it should be for a massless particle that
has only helicity components. We kept $d-1$ components in $\mathbf{\psi }^i$
instead of $d-2$ components that would have been possible by taking the
gauge $\mathbf{\psi \cdot p}=0$. The reason is to maintain manifest rotation
symmetry SO$\left( d-1\right) $, and for this we paid the price of having
the projector $\delta _{ij}-\frac{\mathbf{p}_i\mathbf{p}_j}{\mathbf{p}^2}$.
This projector appears in the anticommutation relations $\left\{ \mathbf{%
\tilde{\psi}}^i,\mathbf{\tilde{\psi}}^j\right\} =\delta _{ij}-\frac{\mathbf{p%
}_i\mathbf{p}_j}{\mathbf{p}^2}$. The generators of SO$\left( d,2\right) $
are obtained by inserting the gauge choice above into the general
expression; obviously the fermions appear only in the form $\tilde{S}^{ij}$.

\subsection{More general potential}

We now discuss another gauge choice, with a rather different embedding of $%
\left( d-1,1\right) $ dimensions in $\left( d+2\right) $ dimensions.
Consider the basis $X^{M}=\left( X^{0^{\prime }},X^{0},X^{I}\right) $ and $%
P^{M}=\left( P^{0^{\prime }},P^{0},P^{I}\right) $ with metric $\eta
^{0^{\prime }0^{\prime }}=\eta ^{00}=-1$ and $\eta ^{IJ}=\delta ^{IJ}$.
Choose one gauge such that the four functions $X^{0^{\prime
}},X^{0},P^{0^{\prime }},P^{0}$ are expressed in terms of three functions $%
F,G,u$ 
\begin{eqnarray}
X^{0^{\prime }} &=&F\cos u,\quad X^{0}=F\sin u\,,  \label{Htt0} \\
P^{0^{\prime }} &=&-G\sin u,\quad P^{0}=G\cos u\,.
\end{eqnarray}%
Inserting this form in the constraints (\ref{constraints}) gives 
\begin{eqnarray}
X^{M} &=&F\left[ \cos u,\,\,\sin u\,\,,\,\,\,n^{I}\right] \,,  \nonumber \\
P^{M} &=&G\left[ -\sin u,\,\cos u\,\,\,,\,\,\,m^{I}\right] \,, \\
\psi ^{M} &=&\,\,\,\,\,\left[ \psi ^{0^{\prime }}\,\,,\quad \psi ^{0}\quad
,\,\,\,\,\psi ^{I}\right] \,,  \nonumber
\end{eqnarray}%
where 
\begin{eqnarray}
\psi ^{0^{\prime }} &=&\cos u\,\,n\cdot \psi -\sin u\,\,m\cdot \psi \,, \\
\psi ^{0} &=&\sin u\,\,n\cdot \psi +\cos u\,\,m\cdot \psi \,,
\end{eqnarray}%
and $n^{I},m^{I}$ are \textit{euclidean} unit vectors that are orthogonal.
We choose the following parametrization for these unit vectors in the basis $%
I=\left[ 1^{\prime },i\right] $ where $I=1^{\prime }$ denotes the extra
space dimension and $i=1,2,\cdots ,(d-1)$ labels ordinary space, 
\begin{eqnarray}
n^{I} &=&\left[ \frac{\sqrt{-2H}}{rV}\,\mathbf{r\cdot p,\quad }(\frac{1}{r}%
\mathbf{r}^{i}+\frac{\mathbf{r\cdot p}}{rV}\mathbf{p}^{i}\mathbf{)}\right] ,
\label{nm} \\
m^{I} &=&\left[ \,\quad (1+\frac{\mathbf{p}^{2}}{V})\quad ,\quad -\frac{%
\sqrt{-2H}}{V}\,\mathbf{p}^{i}\mathbf{\,\,\,\,}\right] ,  \nonumber
\end{eqnarray}%
where 
\begin{equation}
H=\frac{\mathbf{p}^{2}}{2}+V,  \label{HatomH}
\end{equation}%
and $V(\mathbf{r,p,\psi })$ is a priori some potential energy function
(which will be restricted below), giving a non-relativistic Hamiltonian. We
emphasize that this is a solution of the constraints (\ref{constraints})
that had taken the form $n^{I}n^{I}=m^{I}m^{I}=1,$ and $m^{I}n^{I}=0.$ We
still have the freedom of choosing two bosonic gauge functions and two
fermionic gauge functions. These gauge choices will be made as needed in the
discussion below. Since all the constraints are explicitly solved, the $%
A^{ij},F^{i}$ terms drop out in the action (\ref{actionS}) and we get 
\begin{eqnarray}
S_{0} &=&\int_{0}^{T}d\tau \,\,\left( \partial _{\tau }X^{M}P^{N}\,+\frac{i}{%
2}\psi ^{M}\partial _{\tau }\psi ^{N}+0+0\right) \eta _{MN}  \nonumber \\
&=&\int_{0}^{T}d\tau \,\,\left[ 
\begin{array}{c}
GF~m^{I}\partial _{\tau }n^{I}+\frac{i}{2}\psi ^{I}\partial _{\tau }\psi ^{I}
\\ 
-\left[ FG+\frac{1}{2i}\left( n\cdot \psi m\cdot \psi -m\cdot \psi n\cdot
\psi \right) \right] \partial _{\tau }u \\ 
-\frac{i}{2}\left( \,\,n\cdot \psi \,\,\partial _{\tau }\left( n\cdot \psi
\right) +m\cdot \psi \,\partial _{\tau }\left( m\cdot \psi \right) \right) 
\end{array}%
\right]   \label{Haction} \\
&=&\int_{0}^{T}d\tau \,\,\left( \mathbf{p}^{i}\partial _{\tau }\mathbf{r}%
^{i}+\frac{i}{2}\mathbf{\psi }^{i}\partial _{\tau }\mathbf{\psi }^{i}\mathbf{%
-}H\right) .  \nonumber
\end{eqnarray}%
In arriving to the last line we need to use 
\begin{equation}
m^{I}\partial _{\tau }n^{I}=-\frac{\sqrt{-2H}}{rV}\left[ \mathbf{r\cdot p\,}%
\partial _{\tau }\left( \ln \sqrt{-2H}\right) -\partial _{\tau }\left( 
\mathbf{r\cdot p}\right) +\mathbf{p}\cdot \partial _{\tau }\mathbf{r}\right]
.
\end{equation}%
which follows from the $m^{I},n^{I}$ given above, as well as further gauge
fixing described below. A total derivative $\partial _{\tau }\Lambda \left( 
\mathbf{r,p,\psi }\right) $ may be dropped in the last line since it does
not contribute to the variational principle. 

Two bosonic and two fermionic gauge choices remain to be made so that $H$
comes out in the form of Eq.(\ref{HatomH}). The permitted gauge choices must
also be consistent with the original equations of motion that follow from
the original gauge invariant action Eq.(\ref{actionS}), namely 
\begin{equation}
\dot{X}^{M}=A^{22}P^{M}+A^{12}X^{M},\;\dot{P}^{M}=-A^{11}X^{M}-A^{12}P^{M},%
\;i\dot{\psi}^{M}=F^{1}X^{M}+F^{2}P^{M}.  \label{original}
\end{equation}%
The equations of motion that follow from the gauge fixed action, with the
given $H$ above, must be consistent with these original ones. This will
restrict the possible potentials $V$ that can emerge within this class of
gauges. 

Before this last point is checked we first continue to be guided by the form
of the desired gauge fixed action and by symmetry principles. The
consistency of the equations of motion is also related to the conservation
of $J^{MN}=L^{MN}+S^{MN}$ since these gauge invariant quantities are
symmetries of the original gauge invariant action, and therefore they must
remain conserved quantities (some of them as nonlinearly realized hidden
symmetries) of the gauge fixed action in Eq.(\ref{Haction}).  Thus, when the
equations of motion that follow from Eq.(\ref{Haction}) are used, the gauged
fixed $J^{MN}$ must be conserved. This is also a guide in determining the
possible potentials $V$ that can emerge within this class of gauges. To
examine this point we consider in particular $J^{0^{\prime }0}$ which takes
the gauge fixed form%
\begin{eqnarray}
J^{0^{\prime }0} &=&X^{0^{\prime }}P^{0}-X^{0}P^{0^{\prime }}+\frac{1}{2i}%
\left( \psi ^{0^{\prime }}\psi ^{0}-\psi ^{0}\psi ^{0^{\prime }}\right)  
\nonumber \\
&=&FG+\frac{1}{2i}\left( n\cdot \psi m\cdot \psi -m\cdot \psi n\cdot \psi
\right) .
\end{eqnarray}%
We note that in the intermediate steps of the computation in Eq.(\ref%
{Haction}) the coefficient of $\partial _{\tau }u$ is precisely the
conserved $J^{0^{\prime }0}.$ So, in making gauge choices it is wise to
insure that this combination ends up being either a numerical constant or a
constant of motion, such as a function of the Hamiltonian, or other
quantities that commute with the Hamiltonian for some given $V$ (e.g.
angular momentum, etc.). For example, $J^{ij}=L^{ij}+S^{ij}$ is conserved
for a rotationally invariant $V,$ but if $V$ is independent of spin, then
both $L^{ij}$ and $S^{ij}$ will be independently conserved quantities in
addition to $H.$ Hence, in proceeding we will assume that one of the gauge
choices is $FG+\frac{i}{2}\left( m\cdot \psi n\cdot \psi -n\cdot \psi m\cdot
\psi \right) =$constant (independent of equations of motion), or constant of
motion (if the equations of motion of the gauge fixed Hamiltonian are used).

We will not go through all possibilities for $V$ in this paper, but we wish
to give at least one example of gauge choices. The following example
corresponds to the H-atom Hamiltonian, describing the spinning electron
bound by the spin independent potential $V=-\alpha /r$. This gauge is given
by 
\begin{equation}
\psi ^{1^{\prime }}=0,\;\;GF=\frac{\alpha }{\sqrt{-2H}},\;\;u=\left( \mathbf{%
r\cdot p}-2\tau H\right) \frac{\sqrt{-2H}}{\alpha }  \label{Hgauge}
\end{equation}%
There remains one more fermionic gauge choice which can be used to simplify
the gauge fixed action to the one given in Eq.(\ref{Haction}) after using
Eq.(\ref{Hgauge}). To check the consistency of this choice, in this gauge we
compute 
\begin{equation}
m\cdot \psi =\frac{\sqrt{-2H}}{V}\,\mathbf{p\cdot \psi ,\;\;}n\cdot \psi =%
\frac{1}{r}\mathbf{r\cdot \psi }+\frac{\mathbf{r\cdot p}}{rV}\mathbf{p\cdot
\psi ,\;}\frac{1}{2i}\left[ n\cdot \psi ,m\cdot \psi \right] =\frac{\sqrt{-2H%
}}{2\alpha }L_{ij}S^{ij}
\end{equation}%
So we see that $J^{0^{\prime }0}=FG+\frac{1}{2i}\left( n\cdot \psi m\cdot
\psi -m\cdot \psi n\cdot \psi \right) =\frac{\alpha }{\sqrt{-2H}}+\frac{1}{2}%
L_{ij}S^{ij}\frac{\sqrt{-2H}}{\alpha }$ is conserved, as expected, since $%
H,L_{ij},S_{ij}$ are separately conserved when the potential is spin
independent. In fact, one can also perform the stronger check that the
original equations of motion in Eq.(\ref{original}), when gauge fixed, are
fully consistent with the equations of motion for the H-atom that follow
from Eq.(\ref{Haction}).

With a similar procedure one can find other gauge choices that lead to other
potentials, including gauges that lead to spin dependent potentials.

\section{Anti-de Sitter gauge}

It is also possible to find gauges that correspond to particles in various
curved spacetimes. We will ignore the fermions to keep it simple. As an
example we consider the AdS spacetime. Consider the basis $X^M=\left(
X^{0^{\prime }},X^{1^{\prime }},X^m\right) $ with $\eta ^{1^{\prime
}1^{\prime }}=-\eta ^{0^{\prime }0^{\prime }}=1,$ and $\eta ^{mn}=$
Minkowski. The Latin letter $m$ denotes vector components in flat space, and
we will reserve the Greek letter $\mu $ for vector components in curved
space. Choose two gauges $X^{1^{\prime }}=1,\,P^{1^{\prime }}=0$, and solve
the two constraints $X^2=X\cdot P=0$, and let $X^{0^{\prime }}\left(
x\right) ,X^m\left( x\right) $ be given $\,$in terms of $x^{\mu \,\,}$ in
curved space 
\begin{eqnarray}
M &=&\left( \quad 0^{\prime }\quad ,\quad \quad \quad \quad \quad 1^{\prime
}\quad \quad \quad ,\quad \,m\quad \right)  \nonumber \\
X^M &=&\left( \pm \sqrt{1+X_m^2\left( x\right) },\quad \quad 1\quad \quad
,\quad X^m\left( x\right) \,\,\,\right)  \label{genAdS} \\
P^M &=&\left( \frac{X^m\left( x\right) \,e_m^{\mu \,}\left( x\right)
\,p_{\mu \,}}{\pm \sqrt{1+X_m^2\left( x\right) }},\quad 0\quad
,\,\,\,e_m^{\mu \,}\left( x\right) \,p_{\mu \,}\right)  \label{genads2}
\end{eqnarray}
Note that $P^2=0$ has not been imposed yet, and there still is one more
bosonic gauge freedom. Here $e_m^{\mu \,}\left( x\right) $ is the inverse of 
$e_{\mu \,\,}^m\left( x\right) $ defined by 
\begin{equation}
e_{\mu \,\,}^m\left( x\right) =\partial _{\mu \,}X^m\left( x\right)
-X^m\left( x\right) \frac{X\left( x\right) \cdot \partial _{\mu \,}X\left(
x\right) }{1+X_m^2\left( x\right) }.  \label{eads}
\end{equation}
It is designed just such that $p_{\mu \,}$ has the meaning of canonical
momentum when we insert this gauge in the action 
\begin{eqnarray}
S_0 &=&\int_0^Td\tau \,\,\left( \partial _{\tau \,}X^MP^N\,\eta _{MN}-\frac
12A^{22}P\cdot P-0-0\right)  \nonumber \\
&=&\int_0^Td\tau \,\left( \dot{x}^{\mu \,}\cdot p_{\mu \,}-\frac
12A^{22}G^{\mu \nu }\left( x\right) \,p_{\mu \,}p_{\nu \,}\right)
\end{eqnarray}
The remaining part of the action imposes the constraint $P^2=0$, 
\begin{equation}
G^{\mu \nu }\left( x\right) \,p_{\mu \,}p_{\nu \,}=0,
\end{equation}
where the inverse metric $G^{\mu \nu }\left( x\right) $ follows from (\ref%
{genads2}) 
\begin{equation}
G^{\mu \nu }=e_m^\mu e_n^\nu \left( \eta ^{mn}-\frac{X^nX^m}{1+X^2}\right) .
\end{equation}
Taking its inverse one finds $G_{\mu \nu }$ 
\begin{equation}
G_{\mu \nu }=e_\mu ^me_\nu ^n\left( \eta _{nm}+X_nX_m\right) =\partial _\mu
X\cdot \partial _\nu X-\frac{\left( X\cdot \partial _\mu X\right) \left(
X\cdot \partial _\nu X\right) }{1+X^2},
\end{equation}
It turns out that this coincides with the metric obtained from the two
conditions $X^2=0$ and $ds^2=dX\cdot dX=dx^\mu dx^\nu G_{\mu \nu }\left(
x\right) $, using any parametrization for $X^{0^{\prime }}\left( x\right)
,X^\mu \left( x\right) $, in the gauge $X^{1{\prime}}=1$ 
\begin{eqnarray}
G_{\mu \nu }\left( x\right) &=&\partial _{\mu \,}X^m\left( x\right) \partial
_{\nu \,}X^n\left( x\right) \eta _{mn}-\partial _{\mu \,}X^{0^{\prime
}}\left( x\right) \partial _{\nu \,}X^{0^{\prime }}\left( x\right) ,
\label{adsG} \\
-1 &=&X^m\left( x\right) X^n\left( x\right) \eta _{mn}-X^{0^{\prime }}\left(
x\right) X^{0^{\prime }}\left( x\right)
\end{eqnarray}
The last form (\ref{adsG}) makes it evident that this is the AdS metric in $%
\left( d-1,1\right) $ dimensions. To construct it explicitly, one may choose
any convenient function for $X^m\left( x\right) ,$ find the corresponding $%
X^{0^{\prime }}\left( x\right) $ and insert it into (\ref{adsG}). See below
for some examples.

In the quantum theory the constraint is imposed on states $|\phi >$. It is
useful to consider the field $\phi \left( x\right) =<x|\phi >$. The
constraint equation becomes a differential equation on this field. The
operators involved in the constraint must be ordered. A natural ordering
corresponds to the Laplacian condition 
\begin{equation}
\frac 1{\sqrt{-G}}\partial _\mu \left( \sqrt{-G}G^{\mu \nu }\left( x\right)
\partial _\nu \phi \left( x\right) \right) =0.
\end{equation}
The effective field theory that gives this equation is 
\begin{equation}
S_{eff}=\frac 12\int d^dx\,\sqrt{-G}\,G^{\mu \nu }\partial _\mu \phi
\partial _\nu \phi .
\end{equation}

We have seen that the OSp$\left( 1/2\right) $ gauge covariant action (\ref%
{actionS}) is capable of describing curved spacetime as well. The underlying
reason for this is the ability to choose time as a gauge in non-unique ways
because we have more than one timelike coordinate in the $d+2$ dimensional
spacetime. For each choice of time embedded in $d+2$ dimensions the
corresponding canonical Hamiltonian looks different. In particular, the
topology and geometry of the embedding in $d+2$ dimensions is different than
the previous cases. Nevertheless these systems are OSp$\left( 1/2\right) $
gauge equivalent, or dual to each other, since they all correspond to the
same action and same representation of SO$\left( d,2\right) .$

\subsection{Case \#1 for $d=2$}

Consider the following AdS parametrization for $d=2$, which solves all the
constraints $X^2=P^2=X\cdot P=X\cdot \psi =P\cdot \psi =0$ (including
fermions) and gives an explicit metric. Here $\varepsilon =sign\left(
p\right) =\pm 1$ is present to insure that the Hamiltonian is positive (see
below). We can still choose two fermionic gauges, such as $\xi =\chi =0$
which make $\psi ^M$ trivial, however we will not choose a fermionic gauge
yet and see that at the end, gauge invariant quantities, such as the
Hamiltonian and the conformal generators, do not depend on these fermions at
the classical level (but they do at the quantum level as we will see). 
\begin{eqnarray}
M &=&\left[ 0^{\prime }\quad ,\,\quad \quad 1^{\prime }\,\,,\,\,\quad \quad
\quad \quad \quad \quad 0\quad \quad \quad \quad ,\quad \quad \quad 1\quad %
\right]  \nonumber \\
X^M &=&\left[ \varepsilon \cosh x\,\cos t,\,\,\,\,\quad 1\,\,\,\,\,\,\,\quad
,\quad \varepsilon \cosh x\,\sin t,\quad \sinh x\right]  \label{case1} \\
P^M &=&\left[ 
\begin{array}{l}
\varepsilon p\sinh x\,\cos t \\ 
-p\sin t%
\end{array}
,\,0\quad ,\,\, 
\begin{array}{l}
\varepsilon p\sinh x\,\sin t \\ 
+p\cos t%
\end{array}
\,,\,p\cosh x\right]  \label{case11} \\
\psi ^M &=&\left[ 
\begin{array}{l}
\varepsilon \xi \cos t \\ 
+\xi \sinh x\sin t \\ 
-\chi \cosh x\sin t%
\end{array}
\,\,,\, 
\begin{array}{l}
\xi \cosh x \\ 
-\chi \sinh x%
\end{array}
\,\,,\, 
\begin{array}{l}
\varepsilon \xi \sin t \\ 
-\xi \sinh x\cos t \\ 
+\chi \cosh x\cos t%
\end{array}
,\,\,\,\chi \right]  \label{case111}
\end{eqnarray}
We need to evaluate the derivatives 
\begin{eqnarray}
dX^M &=&\left[ 
\begin{array}{c}
-\varepsilon dt\cosh x\,\sin t \\ 
+\varepsilon p\,dx\sinh x\,\cos t%
\end{array}
,0\,, 
\begin{array}{c}
\varepsilon dt\cosh x\,\cos t \\ 
+\varepsilon dx\sinh x\,\sin t%
\end{array}
,\,dx\cosh x\right] \\
d\psi ^M|_{t=0} &=&\left[ 
\begin{array}{c}
\xi dt\sinh x \\ 
-\chi dt\cosh x%
\end{array}
, 
\begin{array}{l}
\xi dx\sinh x \\ 
-\chi dx\cosh x%
\end{array}
, 
\begin{array}{c}
\varepsilon \xi dt-\xi dx\cosh x \\ 
+\chi dx\sinh x%
\end{array}
,0\right]  \nonumber \\
&&+\left[ \varepsilon d\xi ,\, 
\begin{array}{l}
d\xi \cosh x \\ 
-d\chi \sinh x%
\end{array}
\,\,,\,\, 
\begin{array}{c}
-d\xi \sinh x \\ 
+d\chi \cosh x%
\end{array}
,\,\,d\chi \right]
\end{eqnarray}
The metric in $\left( t,x\right) $ space is obtained by computing 
\begin{eqnarray}
ds^2 &=&\left( dX\right) ^2=-dt^2\cosh ^2x+\left( dx\right) ^2 \\
\psi \cdot d\psi &=&\left( \xi \chi -\chi \xi \right) \left( dt\,\varepsilon
\cosh x-dx\right)
\end{eqnarray}
Although we have used $d\psi ^M|_{t=0}$ in this computation for convenience,
the result is valid for any $t$. This form also gives the Lagrangian in the
second order form 
\begin{eqnarray}
L &=&\frac 1{2A^{22}}\left( \partial _\tau X\right) ^2+\frac i2\psi \cdot
\partial _\tau \psi  \nonumber \\
&=&\frac 1{2A^{22}}\left[ -\left( \partial _\tau t\right) ^2\cosh ^2x+\left(
\partial _\tau x\right) ^2\right] \\
&&+s\,\left( -\partial _\tau t\,\varepsilon \cosh x+\partial _\tau x\right) 
\nonumber
\end{eqnarray}
where 
\begin{equation}
s=\frac 1{2i}\left( \xi \chi -\chi \xi \right) .
\end{equation}
Note that there is no kinetic term for the fermions $\xi ,\chi $ hence they
are not dynamical, and we will see that they are just gauge freedom.
Alternatively, after making the gauge choice $t\left( \tau \right) =\tau $,
the Lagrangian in the first order formalism is given directly by (\ref%
{actionS}) 
\begin{eqnarray}
L &=&\partial _\tau X\cdot P+\frac i2\psi \cdot \partial _\tau \psi +0\,\,+0
\\
&=&\dot{x}\left( p+s\right) -H,  \nonumber
\end{eqnarray}
where 
\begin{equation}
H=\left( p+s\right) \varepsilon \cosh x=\left| p+s\right| \cosh x.
\end{equation}
The true canonical momentum is $P=p+s$, and to insure positivity of the
Hamiltonian we choose $\varepsilon $ to be 
\begin{equation}
\varepsilon =sign\left( p+s\right)
\end{equation}
We see that $s$ is completely absorbed into the definition of the canonical
momentum and it disappeared from the gauge invariant Hamiltonian. This means
that we could have taken $s=0$ from the beginning as a gauge choice, thus
preserving the definition of the canonical momentum as $P=p$.

The SO$\left( 2,2\right) $ generators are evaluated by inserting the gauge
choice into the general expression at $\tau =t=0$. If $s$ is allowed from
the beginning we find that it appears everywhere in the combination $p+s.$
So, we set $s=0$ as a gauge choice. The result is 
\begin{eqnarray}
J^{0^{\prime }0} &=&\left| p\right| \cosh x,\quad J^{0^{\prime }1^{\prime
}}=-\left| p\right| \sinh x,\quad J^{0^{\prime }1}=\left| p\right| \\
J^{1^{\prime }1} &=&\,p\cosh x,\quad \,\,\,\,\,\,\,\,\,\,J^{01}=-p\sinh
x,\quad \,\,\,\,\,\,\,\,J^{1^{\prime }0}=p
\end{eqnarray}
These satisfy the SO$\left( 2,2\right) $ algebra at the classical level. By
taking linear combinations we may construct the SO$\left( 2,2\right) =$SL$%
\left( 2,R\right) _L\otimes SL\left( 2,R\right) _R$ generators $%
J_{0,1,2}^{L,R}$ in the following form 
\begin{eqnarray}
J_0^L\pm J_1^L &=&\frac 12\left( p-\left| p\right| \right) e^{\mp x},\quad
J_2^L=\frac 12\left( p-\left| p\right| \right) \\
J_0^R\pm J_1^R &=&\frac 12\left( p+\left| p\right| \right) e^{\mp x}\,,\quad
J_2^R=\frac 12\left( p+\left| p\right| \right)
\end{eqnarray}
We see that either the left moving or the right moving generators must
vanish in momentum space (but not in x-space or other quantum space). The
quadratic Casimirs for both SL$\left( 2,R\right) _{L,R}$ vanish at the
classical level 
\begin{equation}
C_2^{L,R}=\left( J_0^{L,R}+J_1^{L,R}\right) \left(
J_0^{L,R}-J_1^{L,R}\right) -\left( J_2^{L,R}\right) ^2=0.
\end{equation}

\subsection{Quantum ordering case \#1}

We now need to order the operators at the quantum level and make sure that
the Casimir operator is consistent with the gauge invariance requirements at
the quantum level. Recall that for the purely bosonic system Sp$\left(
2\right) $ gauge invariance we must have $C_2\left( SO\left( d,2\right)
\right) =1-d^2/4$ and for the OSp$\left( 1/2\right) $ gauge invariance we
must have $C_2\left( SO\left( d,2\right) \right) =-\frac 18\left( d+2\right)
\left( d-1\right) $. For our case $d=2$ we must have $C_2\left( SO\left(
2,2\right) \right) =0$ for the purely bosonic and $C_2\left( SO\left(
2,2\right) \right) =-1/2$ for the fermionic cases. We see that the fermions
must play a role.

Let us first deal with the purely bosonic case. For either the left or right
movers we need hermitian generators. There is ambiguity in the quantum
ordering as illustrated by the following possible hermitian quantum ordering
of the classical $e^xp$%
\begin{equation}
e^{x/2}pe^{x/2},\quad p^{1/2}e^xp^{1/2},\quad p^\lambda e^{x/2}p^{1-2\lambda
}e^{x/2}p^\lambda ,\quad \cdots
\end{equation}
and similarly for $e^{-x}p$ ordering.. If one re-orders these to the first
form we find 
\begin{eqnarray}
e^xp &\rightarrow &p^\lambda e^{x/2}p^{1-2\lambda }e^{x/2}p^\lambda
=e^{x/2}\left( p^2+\frac 14\right) ^\lambda p^{1-2\lambda }e^{x/2} \\
e^{-x}p &\rightarrow &p^{\lambda ^{^{\prime }}}e^{-x/2}p^{1-2\lambda
^{^{\prime }}}e^{-x/2}p^{\lambda ^{^{\prime }}}=e^{-x/2}\left( p^2+\frac
14\right) ^{\lambda ^{^{\prime }}}p^{1-2\lambda ^{^{\prime }}}e^{-x/2}
\end{eqnarray}
We may also take $\lambda ,\lambda ^{\prime }$ different from each other. In
fact we find that as long as 
\begin{equation}
\lambda +\lambda ^{\prime }=1
\end{equation}
the quantum ordered generators close correctly and they give the Casimir $%
C_2=0$. Thus, let us take $\lambda =\frac 12+\alpha $ and $\lambda ^{\prime
}=\frac 12-\alpha $. Then we have 
\begin{equation}
J_0\pm J_1=e^{\mp x/2}\left( p^2+\frac 14\right) ^{\frac 12\mp \alpha
}p^{\pm 2\alpha }\,e^{\mp x/2},\quad J_2=p
\end{equation}
with commutation rules 
\begin{eqnarray}
\left[ J_0+J_1,J_0-J_1\right] &=&e^{-x/2}\left( p^2+\frac 14\right)
\,e^{x/2}-e^{x/2}\left( p^2+\frac 14\right) \,e^{-x/2}  \nonumber \\
&=&\left( p-\frac i2\right) ^2-\left( p+\frac i2\right) ^2=-2ip \\
&=&-2iJ_2  \nonumber
\end{eqnarray}
and Casimir 
\begin{eqnarray}
C\left( SL\left( 2,R\right) \right) &=&\frac 12\left( J_0+J_1\right) \left(
J_0-J_1\right) +\frac 12\left( J_0-J_1\right) \left( J_0+J_1\right) -\left(
J_2\right) ^2  \nonumber \\
&=&\frac 12e^{-x/2}\left( p^2+\frac 14\right) \,e^{x/2}+\frac
12e^{x/2}\left( p^2+\frac 14\right) \,e^{-x/2}-p^2  \nonumber \\
&=&\frac 12\left( p-\frac i2\right) ^2+\frac 18+\frac 12\left( p+\frac
i2\right) ^2+\frac 18-p^2 \\
&=&0  \nonumber
\end{eqnarray}
In particular the values $\alpha =0,1/2$ yield interesting looking
generators 
\begin{eqnarray}
\alpha &=&0:\quad \,\,\,J_0\pm J_1=e^{\mp x/2}\left( p^2+\frac 14\right)
^{\frac 12}e^{\mp x/2},\quad J_2=p \\
\alpha &=&1/2:\quad \left\{ 
\begin{array}{l}
J_0+J_1=e^{-x/2}pe^{-x/2},\quad J_2=p \\ 
J_0-J_1=e^{x/2}\left( p+\frac 1{4p}\right) \,e^{x/2}%
\end{array}
\right.
\end{eqnarray}
There is no way to decide which of these versions one should use for our
problem.

Next we return to the spinning case. The following modification of the
purely bosonic generators give the desired result for the $\alpha =0,1/2$
cases 
\begin{eqnarray}
\alpha &=&0:\quad \,\,\,J_0\pm J_1=e^{\mp x/2}\left[ \left( p^2+\frac
14\right) ^{\frac 12}\pm \gamma \right] e^{\mp x/2},\quad J_2=p \\
\alpha &=&1/2:\quad \left\{ 
\begin{array}{c}
J_0+J_1=e^{-x/2}pe^{-x/2},\quad J_2=p+\gamma \\ 
J_0-J_1=e^{x/2}\left( p+\frac 1{4p}+2\gamma \right) \,e^{x/2}%
\end{array}
\right.
\end{eqnarray}
Then the SL$\left( 2,R\right) $ algebra closes correctly and the Casimir is 
\begin{equation}
C_2=-\gamma ^2
\end{equation}
The choice $\gamma ^2=1/2$ matches the spinning case.

It may be of interest to note the following more general construction of SL$%
\left( 2,R\right) $. Instead of the form parametrized by $\alpha $ we can
use a more general function $F(p)$%
\begin{equation}
J_0\pm J_1=e^{\mp x/2}\left[ \left( p^2+\frac 14\right) ^{\frac 12}\right]
F^{\pm 1}\,e^{\mp x/2},\quad J_2=p
\end{equation}
with Casimir $C_2=0.$ Some choices of $F(p)$ are interesting. For example,
taking $F=\left( p^2+\frac 14\right) ^{-\frac 12}$ yields 
\begin{equation}
J_0+J_1=e^{-x},\quad J_0-J_1=e^{x/2}\left( p^2+\frac 14\right)
\,e^{x/2},\quad J_2=p.  \label{newads}
\end{equation}
It is interesting to note that we can find a gauge choice for $X^M,P^M$ that
yields these SO$\left( 2,2\right) \,$generators at $\tau =0$, namely, in (%
\ref{case1}-\ref{case111}) replace everywhere $\cosh x$ by $c(x,p)=\frac
1{2p}\left( p^2e^x+e^{-x}\right) $, and $\sinh x$ by $s(x,p)=\frac
1{2p}\left( p^2e^x+e^{-x}\right) $ and then proceed the same way. Since $%
c^2(x,p)-s^2(x,p)=1$, the computation produces similar expressions, ending
with the form (\ref{newads}). Finally, this may be modified with a parameter 
$\gamma $ 
\begin{equation}
J_0+J_1=e^{-x},\quad J_0-J_1=e^{x/2} \left ( p^2+\frac 14-{\gamma}^2 \right
) \,e^{x/2},\quad J_2=p
\end{equation}
to yield a Casimir $C_2=-\gamma ^2$.

\subsection{Case \#2 for d=2}

Consider the following AdS parametrization for $d=2$, which solves all the
constraints. By using similar methods to case \#1 we compute the metric,
action and SO$\left( 2,2\right) $ generators. 
\begin{eqnarray}
M &=&\left[ 0^{\prime }\quad ,\,\quad \quad 1^{\prime }\,\,,\,\,\quad \quad
\quad \quad \quad \quad 0\quad \quad \quad \quad ,\quad \quad \quad 1\quad %
\right]  \nonumber \\
X^M &=&\left[ -\csc x\,\cos t,\,\,\,\,\quad 1\,\,\,\,\,\,\,\quad ,\quad
-\csc x\,\sin t,\quad -\cot x\right] \\
P^M &=&\left[ 
\begin{array}{l}
\left| p\right| \sin x\,\sin t \\ 
+p\cos x\cos t%
\end{array}
,\,\quad 0\quad ,\,\, 
\begin{array}{l}
-\left| p\right| \sin x\,\cos t \\ 
+p\cos x\sin t%
\end{array}
\,,\,\quad p\right] \\
\psi ^M &=&\xi X^M+\chi \frac{P^M}p\,
\end{eqnarray}
The metric is 
\begin{eqnarray}
ds^2 &=&\left( dX\right) ^2=\,\frac 1{\sin ^2x}\,\left( -dt^2+dx^2\right) \\
\psi \cdot d\psi &=&s\left( -\varepsilon dt\,+dx\right)
\end{eqnarray}
where $s=\frac 1{2i}\left( \xi \chi -\chi \xi \right) .$ The quantity $s$ is
absorbed into the definition of true canonical momentum and it disappears.
Thus we take it $s=0$ from the beginning, and compute the Lagrangian 
\begin{eqnarray}
L &=&\partial _\tau X\cdot P+\frac i2\psi \cdot \partial _\tau \psi +0\,\,+0
\nonumber \\
&=&\dot{x}p-H,\quad with\quad H=\left| p\right| .
\end{eqnarray}
The SO$\left( 2,2\right) $ generators are 
\begin{eqnarray}
J^{0^{\prime }0} &=&\left| p\right| ,\quad J^{1^{\prime }0^{\prime }}=p\cos
x,\quad \,\,\,\,J^{10^{\prime }}=p\sin x \\
J^{1^{\prime }1} &=&\,p,\quad \,\,\,\,\,\,\,J^{10}=\left| p\right| \cos
x,\quad J^{01^{\prime }}=\left| p\right| \sin x
\end{eqnarray}
These satisfy the SO$\left( 2,2\right) $ algebra at the classical level. The
SO$\left( 2,2\right) =$SL$\left( 2,R\right) _L\otimes SL\left( 2,R\right) _R$
generators $J_{0,1,2}^{L,R}$ are 
\begin{eqnarray}
J_0^R &=&\frac 12\left( \left| p\right| +p\right) ,\quad J_1^R\pm
iJ_2^R=\frac 12\left( \left| p\right| +p\right) e^{\pm ix}\,, \\
J_0^L &=&\frac 12\left( \left| p\right| -p\right) ,\quad J_1^L\pm
iJ_2^L=\frac 12\left( \left| p\right| -p\right) e^{\pm ix}\,.
\end{eqnarray}
We see that either the left moving or the right moving generators must
vanish in momentum space (but not in x-space or other quantum space). The
quadratic Casimir for both SL$\left( 2,R\right) _{L,R}$ vanishes at the
classical level 
\begin{equation}
C_2^{L,R}=\left( J_0^{L,R}\right) ^2-\left( J_i^{L,R}+iJ_2^{L,R}\right)
\left( J_1^{L,R}-iJ_2^{L,R}\right) =0
\end{equation}

\subsection{Quantum ordering case \#2}

We now need to order the operators at the quantum level and make sure that
the Casimir operator is consistent with the gauge invariance requirements at
the quantum level. For the bosonic case we must have $C_2\left( SO\left(
d,2\right) \right) =1-d^2/4=0$ and with fermions we must have $C_2\left(
SO\left( d,2\right) \right) =-\frac 18\left( d+2\right) \left( d-1\right)
=-1/2$. Let us first deal with the purely bosonic case. For either the left
or right movers we need hermitian generators $J_{1,2,0}^{L,R}$, which
implies $J_1^{L,R}-iJ_2^{L,R}=\left( J_1^{L,R}+iJ_2^{L,R}\right) ^{\dagger }$%
. The following ordering of operators is hermitian for any real number $%
\alpha $ 
\begin{equation}
J_1\pm iJ_2=e^{\pm ix/2}\left( p^2-\frac 14\right) ^{\frac 12\mp \alpha
}p^{\pm 2\alpha }\,e^{\pm ix/2},\quad J_0=p.
\end{equation}
The commutation rules close for any $\alpha $ 
\begin{eqnarray}
\left[ J_1+iJ_2,J_1-iJ_2\right] &=&e^{ix/2}\left( p^2-\frac 14\right)
\,e^{-ix/2}-e^{-ix/2}\left( p^2-\frac 14\right) \,e^{ix/2}  \nonumber \\
&=&\left( p-\frac 12\right) ^2-\left( p+\frac 12\right) ^2=-2p \\
&=&-2J_0  \nonumber
\end{eqnarray}
and the Casimir is zero 
\begin{eqnarray}
C\left( SL\left( 2,R\right) \right) &=&\left( J_0\right) ^2-\frac 12\left(
J_1+iJ_2\right) \left( J_1-iJ_2\right) -\frac 12\left( J_1-iJ_2\right)
\left( J_1+iJ_2\right)  \nonumber \\
&=&p^2-\frac 12e^{ix/2}\left( p^2-\frac 14\right) \,e^{-ix/2}-\frac
12e^{-ix/2}\left( p^2-\frac 14\right) \,e^{ix/2}  \nonumber \\
&=&p^2-\frac 12\left( p-\frac 12\right) ^2+\frac 18-\frac 12\left( p+\frac
12\right) ^2+\frac 18 \\
&=&0.  \nonumber
\end{eqnarray}
In particular the value $\alpha =0$ yields the generators 
\begin{equation}
\alpha =0:\quad \,\,\,J_1\pm iJ_2=e^{\pm ix/2}\left( p^2-\frac 14\right)
^{\frac 12}e^{\pm ix/2},\quad J_0=p
\end{equation}
There is no way to decide which of these $\alpha $ versions one should use
for our problem.

Next we return to the spinning case. The following modification of the
purely bosonic generators give the desired result for the $\alpha =0$ case 
\begin{equation}
\alpha =0:\quad \,\,\,J_1\pm iJ_2=e^{\pm ix/2}\left[ \left( p^2-\frac
14\right) ^{\frac 12}\pm \gamma \right] e^{\pm ix/2},\quad J_0=p
\end{equation}
Then the SL$\left( 2,R\right) $ algebra closes correctly and the Casimir is 
\begin{equation}
C_2=-\gamma ^2
\end{equation}
The choice $\gamma ^2=1/2$ matches the spinning case.

In addition, there are also other orderings, such as 
\begin{eqnarray}
J_1+iJ_2 &=&\left( p+\alpha \right) e^{ix}=\left( p+\alpha +1\right) e^{ix}
\\
J_1-iJ_2 &=&e^{-ix}\left( p+\alpha ^{*}\right) =e^{-ix}\left( p+\alpha
^{*}+1\right) \\
J_0 &=&p+\frac 12\left( 1+\alpha +\alpha ^{*}\right)
\end{eqnarray}
where $\alpha $ is a complex number to be determined by fixing the Casimir
operator. The algebra closes and the Casimir is 
\begin{equation}
C_2=\frac 14\left( -1+\alpha ^2+\alpha ^{*2}-2\left| \alpha \right|
^2\right) .  \label{scasimir}
\end{equation}
We may choose many possible values for $\alpha $ (e.g. $\alpha =(\cot \theta
+i)/\sqrt{8}$) so that $C_2=-1/2.$

\subsection{General $d$}

As an example for general $d$ consider the following choice of AdS gauge
consistent with the general formula (\ref{genAdS}), parametrized in terms of 
$x^\mu =\left( t,\mathbf{r}\right) $ and $p^\mu =\left( H,\mathbf{p}\right) $
\begin{eqnarray}
M &=&\left[ 0^{\prime },1^{\prime },0,i\right]  \nonumber \\
X^M &=&\left[ \frac{r^2+1}{2r}\cos t,\,\,\,1\,\,\,\,,\,\,\frac{r^2+1}{2r}%
\sin t\,,\,\,\frac{r^2-1}{2r^2}\,\mathbf{r}\right] \\
P^M &=&\left[ 
\begin{array}{l}
\frac{r^2-1}{2r}\,\mathbf{r\cdot p}\cos t \\ 
-\frac{2rH}{r^2+1}\sin t%
\end{array}
,\,0\,\,,\,\, 
\begin{array}{l}
\frac{r^2-1}{2r}\,\mathbf{r\cdot p}\sin t \\ 
+\frac{2rH}{r^2+1}\cos t%
\end{array}
\,\,,\,\, 
\begin{array}{l}
\frac{2r^2}{r^2-1}(\mathbf{p-}\frac{\mathbf{r}}{r^2}\mathbf{r\cdot p)} \\ 
+\frac{r^2+1}{2r^2}\,\mathbf{r\cdot p\,\,r}%
\end{array}
\right]
\end{eqnarray}
The metric $G_{\mu \nu }$ is given by 
\begin{equation}
ds^2=dX\cdot dX=-\left( \frac{r^2+1}{2r}\right) ^2dt^2+\frac
1{r^2}dr^2+\left( \frac{r^2-1}{2r}\right) ^2\left( d\Omega \right) ^2{.}
\end{equation}
The classical Hamiltonian $H=p^0$ follows from $P^2=G_{\mu \nu }p^\mu p^\nu
=0$ 
\begin{equation}
H=\frac{r^2+1}2\sqrt{p_r^2+\left( \frac 2{r^2-1}\right) ^2\frac
12L^{ij}L_{ij}}.
\end{equation}
The SO$\left( d,2\right) $ generators $L^{MN}=X^MP^N-X^NP^M$ may now be
constructed by inserting the gauge choice at $t=0$. The form is complicated
and operator quantum ordering is difficult in this gauge. Therefore we will
not go into details.

\section{Conformal gauge}

The particle gauge in (\ref{relativistic}) may be modified by an overall
multiplicative function $F\left( x\right) $ 
\begin{eqnarray}
M &=&\left[ \quad +^{\prime }\quad ,\quad -^{\prime }\quad ,\quad \mu \quad %
\right] \,,  \nonumber \\
X^M &=&\left[ \quad 1\quad ,\quad x^2/2\quad ,\quad x^\mu \quad \right]
\,F\left( x\right) , \\
P^M &=&\left[ \quad 0\quad ,\quad x\cdot p\quad ,\quad p^\mu \quad \right]
\,\frac 1{F\left( x\right) },
\end{eqnarray}
The $P^2=0$ constraint is yet to be imposed. The methods are similar to
those used for the AdS gauge. The metric that corresponds to this gauge
choice is 
\begin{equation}
ds^2=dX^MdX_M=F^2dx^\mu dx_\mu .
\end{equation}
Therefore $F^2$ plays the role of the conformal factor for an arbitrary
conformal metric in $d$ dimensions 
\begin{equation}
G_{\mu \nu }=F^2\left( x\right) \eta _{\mu \nu }.
\end{equation}
The momentum constraint takes the form 
\begin{equation}
P^2=G^{\mu \nu }p_\mu p_\nu =\frac{\eta ^{\mu \nu }}{F^2\left( x\right) }%
p_\mu p_\nu =0.
\end{equation}
This is seen also by inserting the gauge into the action (\ref{actionS}),
which becomes (in the absence of fermions) 
\begin{equation}
S=\int d\tau \,\left( \partial _\tau x^\mu p_\mu -\frac 12A^{22}\frac{\eta
^{\mu \nu }}{F^2\left( x\right) }p_\mu p_\nu \right) .
\end{equation}
The quantum ordered version of the constraint is applied on states $|\phi >$
or $\phi \left( x\right) =<x|\phi >$. A good guess is that the quantum
ordering should correspond to the Laplacian for the metric $G_{\mu \nu }$ 
\begin{equation}
\frac 1{\sqrt{-G}}\partial _\mu \left( \sqrt{-G}G^{\mu \nu }\partial _\nu
\phi \left( x\right) \right) =\frac 1{F^d}\partial _\mu \left( F^{d-2}\eta
^{\mu \nu }\partial _\nu \phi \left( x\right) \right) =0.
\end{equation}
The effective field theory that gives this equation is (for the spinless Sp$%
\left( 2\right) $ gauge theory) 
\begin{equation}
S_{eff}=\int d^dx\,F^{d-2}\left( x\right) \partial _\mu \bar{\phi}\partial
_\nu \phi \eta ^{\mu \nu }.
\end{equation}
This is modified to a Dirac equation for the theory with spin (OSp$\left(
1/2\right) $ gauge theory) 
\begin{equation}
S_{eff}=\int d^dx\,F^{d-1}\left( x\right) \overline{\Psi }\gamma ^\mu
\partial ^\nu \Psi \eta _{\mu \nu }.
\end{equation}

The SO$\left( d,2\right) $ generators $J^{MN}=X^MP^N-X^NP^M+S^{MN}$ are
unaltered at the classical level since the $F\left( x\right) $ factor
cancels. However, at the quantum level, they need to be quantum ordered so
that they are hermitian according to the dot product in curved backgrounds 
\begin{eqnarray}
&<&\phi |\phi >=\frac 12\int d^{d-1}x\,F^{d-2}\,\left( \bar{\phi}i\partial
_0\phi -i\partial _0\bar{\phi}\phi \right) , \\
&<&\Psi |\Psi >=\int d^{d-1}x\,F^{d-1}\,\overline{\Psi }\gamma _0\Psi \,.
\end{eqnarray}
After the quantum ordering one should check the Casimir $C_2\left( SO\left(
d,2\right) \right) $ and verify that it is consistent with (\ref{c2sod2})
for the fermionic theory, and with $C_2=1-d^2/4$ for the bosonic theory, as
follows.

To find the correct order of operators consider the condition $<J^{+^{\prime
}-^{\prime }}\phi |\phi >=$ $<\phi |J^{+^{\prime }-^{\prime }}\phi >$ or $%
<J^{+^{\prime }-^{\prime }}\Psi |\Psi >=<\Psi |J^{+^{\prime }-^{\prime
}}\Psi >$. We find that we must have 
\begin{equation}
J^{+^{\prime }-^{\prime }}=\frac 12\left( x\cdot p+p\cdot x\right)
+is_0+\frac i2\left( d-2\right) \,x\cdot \partial \ln F\left( x\right) .
\end{equation}
The first term $\frac 12\left( x\cdot p+p\cdot x\right) $ is the hermitian
ordering for a dot product with naive integration measure. The quantum
correction $is_0$, was already present in flat space due to hermiticity with
a more involved dot product ($s_0=1$ for $\phi $, and $s_0=1/2$ for $\Psi $;
see \cite{dualconf} and eq.(\ref{ConfLorentz})). The last term is required
in the conformal curved background $F$ with the dot products given above.
The proof of hermiticity uses the conservation of the current $J^\mu =\frac
i2F^{d-2}\left( \bar{\phi}\partial ^\mu \phi -\partial ^\mu \bar{\phi}\phi
\right) $ or $F^{d-1}\overline{\Psi }\gamma ^\mu \Psi ,$ i.e. $\partial _\mu
J^\mu =0,$ that follows from the equation of motion (i.e. constraint). This
expression for $J^{+^{\prime }-^{\prime }}$ may be rewritten in the form 
\begin{equation}
J^{+^{\prime }-^{\prime }}=\frac 12\left( x\cdot \tilde{p}+\tilde{p}\cdot
x\right) +is_0\,,
\end{equation}
where $\tilde{p}^\mu $ is the following order of operators 
\begin{equation}
\tilde{p}^\mu =F^{\frac 12\left( d-2\right) }\left( x\right) \,p^\mu
\,F^{-\frac 12\left( d-2\right) }\left( x\right) =p^\mu +i\left( d-2\right)
\,x\cdot \partial \ln F\left( x\right) .
\end{equation}
We find that the rest of the generators $J^{MN}$ are also hermitian provided
we use the result for flat space (\ref{ConfLorentz}) and replace everywhere $%
p^\mu \rightarrow \tilde{p}^\mu $. The generators in curved conformal space
are then given in terms of those in flat space by the prescription 
\begin{equation}
J_{conf}^{MN}(x,\tilde{p})=F^{\frac 12\left( d-2\right) }\left( x\right)
\,J_{flat}^{MN}(x,p)\,\,F^{-\frac 12\left( d-2\right) }\left( x\right) .
\end{equation}
Then the Casimir operator becomes 
\begin{eqnarray}
C_2\left( SO\left( d,2\right) \right) _{conf} &=&F^{\frac 12\left(
d-2\right) }\left( x\right) \,C_2\left( SO\left( d,2\right) \right)
_{flat}\,\,F^{-\frac 12\left( d-2\right) }\left( x\right) \\
&=&C_2\left( SO\left( d,2\right) \right) _{flat}\,,  \nonumber
\end{eqnarray}
where the last step holds since $C_2\left( SO\left( d,2\right) \right)
_{flat}$ is independent of $x$ or $p$ (see (\ref{C2Lorentz})). The same is
true for all higher Casimir operators because the orbital part $x,p$ drops
out \cite{hermsymm}. This proves again that the quantum theory in the
conformal gauge has the same quantum Hilbert space as all other gauges.

Similarly, one may go through all the previous gauges that are closely
associated with the particle gauge. These include the lightcone gauge (\ref%
{relparticl}) and the timelike gauge (\ref{timelike}). It would be
interesting to study the modifications in the presence of the conformal
factor $F$ for these cases since this generates new representations of SO$%
\left( d,2\right) $ for each choice of $F$.

\section{OSp$\left( n/2\right) $ gauge theory and higher spins}

We can generalize the OSp$\left( 1/2\right) $ theory by adding more copies
of the fermions $\psi _a^M$ with $a=1,2,\cdots ,n$. This provides the
possibility of describing particles and other systems with higher spins.
Thus, consider the fundamental representation $\Phi _I^M=\left( \psi
_a^M,X_i^M\right) $ of OSp($n/2$), with $X_1^M=X^M$ and $X_2^M=P^M$ as
before. \bigskip Introduce the gauge fields 
\begin{equation}
A^{IJ}=\left( 
\begin{array}{ll}
B^{\left[ ab\right] } & F^{ai} \\ 
\varepsilon _{ij}F^{jb} & A^{ij}%
\end{array}
\right) ,\quad A,B=bose,\quad F=fermi
\end{equation}
where $B^{\left[ ab\right] }$ is the antisymmetric SO$\left( n\right) $
gauge field and $A^{ij}$is the symmetric Sp$\left( 2\right) $ gauge field,
as before. There are also $2n$ fermionic gauge fields $F^{ai}$. The local
OSp(n/2) gauge invariant Lagrangian is 
\begin{eqnarray}
S_0 &=&\frac 12\int_0^Td\tau \,\,\left( D_\tau \Phi _I^M\right) g^{IJ}\Phi
_J^N\,\,\eta _{MN},\quad g^{IJ}:OSp\,\,metric  \nonumber \\
&=&\int_0^Td\tau \,\,\left[ 
\begin{array}{c}
X_2\cdot \partial _{\tau \,}X_1+\frac i2\psi _a\cdot \partial _{\tau \,}\psi
_a-\frac 12A^{ij}X_i\cdot X_j \\ 
+iF^{ia}X_i\cdot \psi _a-\frac 12B^{ab}\psi _a\cdot \psi _b%
\end{array}
\right] .
\end{eqnarray}
As before, the constraints have non-trivial solutions provided there are two
times, and the global symmetry is SO$\left( d,2\right) $.

Covariant quantization can be carried out as before. In order to have OSp$%
\left( n/2\right) $ singlets all of its Casimirs must vanish. Then we find
that the quadratic Casimir of SO$\left( d,2\right) $ must have the special
value 
\begin{equation}
C_2\left( SO\left( d,2\right) \right) =\frac 18\left( n-2\right) \left(
d+2\right) \left( d+n-2\right) .  \label{Cospn2}
\end{equation}
This is consistent with the $n=0$ case of \cite{dualconf}\cite{dualH} and
the $n=1$ case treated in this paper.

The SO$\left( d-1,1\right) $ Lorentz covariant particle gauge is easy to
analyze 
\begin{eqnarray}
M &=&\left( +^{\prime },\quad -^{\prime },\quad \quad \mu \quad \right) 
\nonumber \\
X^M &=&\left( 1,\quad x^2/2\quad ,\quad x^{\mu \,\quad }\right) \\
P^M &=&\left( 0,\quad x\cdot p\quad ,\quad p^{\mu \,}\quad \right) \quad
p^2=0 \\
\psi _a^M &=&\left( 0,\quad x\cdot \psi _a\quad ,\quad \psi _a^{\mu
\,}\right) \quad p\cdot \psi _a=0=\psi _{[a}\cdot \psi _{b]}
\end{eqnarray}
The remaining constraints and gauge symmetries are those of worldline
supergravity with $n$ supersymmetries. These were studied in \cite{howe}.
From the analysis in \cite{howe} and \cite{siegel} we know that this system
describes massless spinning particles. The effective fields that represent
them are the analogs of gauge fields, i.e. forms that couple to $p$-branes
(with $p=n/2-1$), and their fermionic generalizations 
\begin{eqnarray}
A_{\mu _1\mu _2\cdots \mu _{n/2}}\left( x\right) ,\quad n &=&even, \\
\Psi _{\alpha \mu _1\mu _2\cdots \mu _{(n-1)/2}}\left( x\right) ,\quad n
&=&odd\,.
\end{eqnarray}
When written in this form, the constraints generate the appropriate field
equations that remove the ghosts and give the correct counting of degrees of
freedom in $d$ dimensions.

The SO$\left( d,2\right) $ generators in the particle gauge have the same
form as (\ref{ConfLorentz}), but with 
\begin{equation}
s^{\mu \nu }=\frac 1{2i}\left( \psi _a^{\mu \,}\psi _a^{\nu \,}-\psi _a^{\nu
\,}\psi _a^{\mu \,}\right) ,\quad \frac 12s^{\mu \nu }s_{\mu \nu }=\frac
n8d\left( d+n-2\right) ,
\end{equation}
in the gauge invariant sector of SO$\left( n\right) $ singlets ($\psi
_{[a}\cdot \psi _{b]}=0$). The quadratic Casimir is given in (\ref{C2Lorentz}%
) 
\begin{equation}
C_2=-\frac{d^2}4+s_0^2+\frac 12s^{\mu \nu }s_{\mu \nu }.
\end{equation}
So, now we need 
\begin{equation}
s_0=\left( 1-\frac n2\right) ,
\end{equation}
in order to agree with the requirements that followed from OSp$\left(
n/2\right) $ gauge invariance given in (\ref{Cospn2}). This value of $s_0$
gives the following dimensions for the fields $A_{\mu _1\mu _2\cdots \mu
_{n/2}}\left( x\right) $, $\Psi _{\alpha \mu _1\mu _2\cdots \mu
_{(n-1)/2}}\left( x\right) $%
\begin{equation}
iJ^{+^{\prime }-^{\prime }}\left( A\,\,or\,\,\Psi \right) =d/2-s_0=\frac
12\left( d+n-2\right) .
\end{equation}
This agrees with the $n=0,1$ cases that we have already studied explicitly
in several forms.

\end{document}